\documentclass{emulateapj}

\newcommand{\chiFTFmin}{5.395}
\newcommand{\chiFTFMCMCmin}{5.396}
\newcommand{\chiFTFMCMCplus}{0.003}
\newcommand{\chiFTFMCMCminus}{0.004}
\newcommand{\chiFTFEXPONENT}{5}
\newcommand{\chiReducedFTFmin}{0.9581}
\newcommand{\chiReducedFTFMCMCmin}{0.9582}
\newcommand{\chiReducedFTFMCMCplus}{0.0005}
\newcommand{\chiReducedFTFMCMCminus}{0.0008}
\newcommand{\chiReducedFTFEXPONENT}{0}
\newcommand{\paramONEFTFmin}{-0.05}
\newcommand{\paramONEFTFMCMCmin}{-0.05}
\newcommand{\paramONEFTFMCMCplus}{0.04}
\newcommand{\paramONEFTFMCMCminus}{0.04}
\newcommand{\paramONEFTFEXPONENT}{0}
\newcommand{\paramTWOFTFmin}{0.82}
\newcommand{\paramTWOFTFMCMCmin}{0.82}
\newcommand{\paramTWOFTFMCMCplus}{0.02}
\newcommand{\paramTWOFTFMCMCminus}{0.02}
\newcommand{\paramTWOFTFEXPONENT}{0}
\newcommand{\paramTHREEFTFmin}{0.46}
\newcommand{\paramTHREEFTFMCMCmin}{0.46}
\newcommand{\paramTHREEFTFMCMCplus}{0.02}
\newcommand{\paramTHREEFTFMCMCminus}{0.02}
\newcommand{\paramTHREEFTFEXPONENT}{0}
\newcommand{\paramFOURFTFmin}{3.1}
\newcommand{\paramFOURFTFMCMCmin}{3.1}
\newcommand{\paramFOURFTFMCMCplus}{0.1}
\newcommand{\paramFOURFTFMCMCminus}{0.2}
\newcommand{\paramFOURFTFEXPONENT}{0}
\newcommand{\paramFIVEFTFmin}{-3.09}
\newcommand{\paramFIVEFTFMCMCmin}{-3.10}
\newcommand{\paramFIVEFTFMCMCplus}{0.06}
\newcommand{\paramFIVEFTFMCMCminus}{0.08}
\newcommand{\paramFIVEFTFEXPONENT}{0}
\newcommand{\paramSIXFTFmin}{0.94}
\newcommand{\paramSIXFTFMCMCmin}{0.94}
\newcommand{\paramSIXFTFMCMCplus}{0.04}
\newcommand{\paramSIXFTFMCMCminus}{0.05}
\newcommand{\paramSIXFTFEXPONENT}{0}
\newcommand{\paramSEVENFTFmin}{4.88}
\newcommand{\paramSEVENFTFMCMCmin}{4.86}
\newcommand{\paramSEVENFTFMCMCplus}{0.01}
\newcommand{\paramSEVENFTFMCMCminus}{0.05}
\newcommand{\paramSEVENFTFEXPONENT}{0}
\newcommand{\paramEIGHTFTFmin}{0.49}
\newcommand{\paramEIGHTFTFMCMCmin}{0.49}
\newcommand{\paramEIGHTFTFMCMCplus}{0.02}
\newcommand{\paramEIGHTFTFMCMCminus}{0.02}
\newcommand{\paramEIGHTFTFEXPONENT}{0}
\newcommand{\paramNINEFTFmin}{4.118}
\newcommand{\paramNINEFTFMCMCmin}{4.118}
\newcommand{\paramNINEFTFMCMCplus}{0.004}
\newcommand{\paramNINEFTFMCMCminus}{0.004}
\newcommand{\paramNINEFTFEXPONENT}{0}
\newcommand{\paramTENFTFmin}{13}
\newcommand{\paramTENFTFMCMCmin}{12}
\newcommand{\paramTENFTFMCMCplus}{4}
\newcommand{\paramTENFTFMCMCminus}{10}
\newcommand{\paramTENFTFEXPONENT}{0}
\newcommand{\paramTENONEFTFmin}{-0.95}
\newcommand{\paramTENONEFTFMCMCmin}{-0.95}
\newcommand{\paramTENONEFTFMCMCplus}{0.02}
\newcommand{\paramTENONEFTFMCMCminus}{0.03}
\newcommand{\paramTENONEFTFEXPONENT}{0}
\newcommand{\paramTENTWOFTFmin}{0.98}
\newcommand{\paramTENTWOFTFMCMCmin}{0.97}
\newcommand{\paramTENTWOFTFMCMCplus}{0.05}
\newcommand{\paramTENTWOFTFMCMCminus}{0.07}
\newcommand{\paramTENTWOFTFEXPONENT}{0}
\newcommand{\paramTENTHREEFTFmin}{54}
\newcommand{\paramTENTHREEFTFMCMCmin}{54}
\newcommand{\paramTENTHREEFTFMCMCplus}{3}
\newcommand{\paramTENTHREEFTFMCMCminus}{2}
\newcommand{\paramTENTHREEFTFEXPONENT}{0}
\newcommand{\paramTENFOURFTFmin}{5.33}
\newcommand{\paramTENFOURFTFMCMCmin}{5.34}
\newcommand{\paramTENFOURFTFMCMCplus}{0.24}
\newcommand{\paramTENFOURFTFMCMCminus}{0.09}
\newcommand{\paramTENFOURFTFEXPONENT}{-10}
\newcommand{\paramTENFIVEFTFmin}{2.1}
\newcommand{\paramTENFIVEFTFMCMCmin}{2.2}
\newcommand{\paramTENFIVEFTFMCMCplus}{0.2}
\newcommand{\paramTENFIVEFTFMCMCminus}{0.1}
\newcommand{\paramTENFIVEFTFEXPONENT}{-10}
\newcommand{\paramTENSIXFTFmin}{112.1}
\newcommand{\paramTENSIXFTFMCMCmin}{112.2}
\newcommand{\paramTENSIXFTFMCMCplus}{0.8}
\newcommand{\paramTENSIXFTFMCMCminus}{1.5}
\newcommand{\paramTENSIXFTFEXPONENT}{0}
\newcommand{\paramTENSEVENFTFmin}{0.049}
\newcommand{\paramTENSEVENFTFMCMCmin}{0.049}
\newcommand{\paramTENSEVENFTFMCMCplus}{0.002}
\newcommand{\paramTENSEVENFTFMCMCminus}{0.002}
\newcommand{\paramTENSEVENFTFEXPONENT}{0}
\newcommand{\paramTENEIGHTFTFmin}{0.054}
\newcommand{\paramTENEIGHTFTFMCMCmin}{0.056}
\newcommand{\paramTENEIGHTFTFMCMCplus}{0.009}
\newcommand{\paramTENEIGHTFTFMCMCminus}{0.010}
\newcommand{\paramTENEIGHTFTFEXPONENT}{0}

\newcommand{\chiSOSmin}{6.664}
\newcommand{\chiSOSMCMCmin}{6.664}
\newcommand{\chiSOSMCMCplus}{0.006}
\newcommand{\chiSOSMCMCminus}{0.002}
\newcommand{\chiSOSEXPONENT}{5}
\newcommand{\chiReducedSOSmin}{1.1833}
\newcommand{\chiReducedSOSMCMCmin}{1.1834}
\newcommand{\chiReducedSOSMCMCplus}{0.0010}
\newcommand{\chiReducedSOSMCMCminus}{0.0003}
\newcommand{\chiReducedSOSEXPONENT}{0}
\newcommand{\paramONESOSmin}{0.11}
\newcommand{\paramONESOSMCMCmin}{0.12}
\newcommand{\paramONESOSMCMCplus}{0.06}
\newcommand{\paramONESOSMCMCminus}{0.06}
\newcommand{\paramONESOSEXPONENT}{0}
\newcommand{\paramTWOSOSmin}{0.84}
\newcommand{\paramTWOSOSMCMCmin}{0.85}
\newcommand{\paramTWOSOSMCMCplus}{0.01}
\newcommand{\paramTWOSOSMCMCminus}{0.02}
\newcommand{\paramTWOSOSEXPONENT}{0}
\newcommand{\paramTHREESOSmin}{0.506}
\newcommand{\paramTHREESOSMCMCmin}{0.505}
\newcommand{\paramTHREESOSMCMCplus}{0.028}
\newcommand{\paramTHREESOSMCMCminus}{0.006}
\newcommand{\paramTHREESOSEXPONENT}{0}
\newcommand{\paramFOURSOSmin}{2.73}
\newcommand{\paramFOURSOSMCMCmin}{2.73}
\newcommand{\paramFOURSOSMCMCplus}{0.12}
\newcommand{\paramFOURSOSMCMCminus}{0.08}
\newcommand{\paramFOURSOSEXPONENT}{0}
\newcommand{\paramFIVESOSmin}{-2.89}
\newcommand{\paramFIVESOSMCMCmin}{-2.89}
\newcommand{\paramFIVESOSMCMCplus}{0.07}
\newcommand{\paramFIVESOSMCMCminus}{0.04}
\newcommand{\paramFIVESOSEXPONENT}{0}
\newcommand{\paramSIXSOSmin}{1.05}
\newcommand{\paramSIXSOSMCMCmin}{1.05}
\newcommand{\paramSIXSOSMCMCplus}{0.04}
\newcommand{\paramSIXSOSMCMCminus}{0.03}
\newcommand{\paramSIXSOSEXPONENT}{0}
\newcommand{\paramSEVENSOSmin}{6.312}
\newcommand{\paramSEVENSOSMCMCmin}{6.312}
\newcommand{\paramSEVENSOSMCMCplus}{0.033}
\newcommand{\paramSEVENSOSMCMCminus}{0.010}
\newcommand{\paramSEVENSOSEXPONENT}{0}
\newcommand{\paramEIGHTSOSmin}{0.464}
\newcommand{\paramEIGHTSOSMCMCmin}{0.465}
\newcommand{\paramEIGHTSOSMCMCplus}{0.024}
\newcommand{\paramEIGHTSOSMCMCminus}{0.008}
\newcommand{\paramEIGHTSOSEXPONENT}{0}
\newcommand{\paramNINESOSmin}{3.871}
\newcommand{\paramNINESOSMCMCmin}{3.871}
\newcommand{\paramNINESOSMCMCplus}{0.003}
\newcommand{\paramNINESOSMCMCminus}{0.003}
\newcommand{\paramNINESOSEXPONENT}{0}
\newcommand{\paramTENSOSmin}{1.6}
\newcommand{\paramTENSOSMCMCmin}{1.6}
\newcommand{\paramTENSOSMCMCplus}{0.3}
\newcommand{\paramTENSOSMCMCminus}{0.3}
\newcommand{\paramTENSOSEXPONENT}{0}
\newcommand{\paramTENONESOSmin}{-1.16}
\newcommand{\paramTENONESOSMCMCmin}{-1.16}
\newcommand{\paramTENONESOSMCMCplus}{0.02}
\newcommand{\paramTENONESOSMCMCminus}{0.03}
\newcommand{\paramTENONESOSEXPONENT}{0}
\newcommand{\paramTENTWOSOSmin}{0.84}
\newcommand{\paramTENTWOSOSMCMCmin}{0.84}
\newcommand{\paramTENTWOSOSMCMCplus}{0.05}
\newcommand{\paramTENTWOSOSMCMCminus}{0.02}
\newcommand{\paramTENTWOSOSEXPONENT}{0}
\newcommand{\paramTENTHREESOSmin}{71}
\newcommand{\paramTENTHREESOSMCMCmin}{71}
\newcommand{\paramTENTHREESOSMCMCplus}{3}
\newcommand{\paramTENTHREESOSMCMCminus}{3}
\newcommand{\paramTENTHREESOSEXPONENT}{0}
\newcommand{\paramTENFOURSOSmin}{5.60}
\newcommand{\paramTENFOURSOSMCMCmin}{5.62}
\newcommand{\paramTENFOURSOSMCMCplus}{0.06}
\newcommand{\paramTENFOURSOSMCMCminus}{0.06}
\newcommand{\paramTENFOURSOSEXPONENT}{-10}
\newcommand{\paramTENFIVESOSmin}{2.0}
\newcommand{\paramTENFIVESOSMCMCmin}{2.0}
\newcommand{\paramTENFIVESOSMCMCplus}{0.2}
\newcommand{\paramTENFIVESOSMCMCminus}{0.1}
\newcommand{\paramTENFIVESOSEXPONENT}{-10}
\newcommand{\paramTENSIXSOSmin}{112.7}
\newcommand{\paramTENSIXSOSMCMCmin}{112.7}
\newcommand{\paramTENSIXSOSMCMCplus}{0.9}
\newcommand{\paramTENSIXSOSMCMCminus}{1.2}
\newcommand{\paramTENSIXSOSEXPONENT}{0}
\newcommand{\paramTENSEVENSOSmin}{0.0489}
\newcommand{\paramTENSEVENSOSMCMCmin}{0.0492}
\newcommand{\paramTENSEVENSOSMCMCplus}{0.0010}
\newcommand{\paramTENSEVENSOSMCMCminus}{0.0012}
\newcommand{\paramTENSEVENSOSEXPONENT}{0}
\newcommand{\paramTENEIGHTSOSmin}{0.051}
\newcommand{\paramTENEIGHTSOSMCMCmin}{0.052}
\newcommand{\paramTENEIGHTSOSMCMCplus}{0.012}
\newcommand{\paramTENEIGHTSOSMCMCminus}{0.005}
\newcommand{\paramTENEIGHTSOSEXPONENT}{0}
\newcommand{\paramTENNINESOSmin}{-9992.0}
\newcommand{\paramTENNINESOSMCMCmin}{-9997.0}
\newcommand{\paramTENNINESOSMCMCplus}{70.4}
\newcommand{\paramTENNINESOSMCMCminus}{20.0}
\newcommand{\paramTENNINESOSEXPONENT}{0}
\newcommand{\paramTENTENSOSmin}{-9993}
\newcommand{\paramTENTENSOSMCMCmin}{-9998}
\newcommand{\paramTENTENSOSMCMCplus}{80}
\newcommand{\paramTENTENSOSMCMCminus}{10}
\newcommand{\paramTENTENSOSEXPONENT}{0}

\newcommand{\chiEOFmin}{6.373}
\newcommand{\chiEOFMCMCmin}{6.373}
\newcommand{\chiEOFMCMCplus}{0.008}
\newcommand{\chiEOFMCMCminus}{0.001}
\newcommand{\chiEOFEXPONENT}{5}
\newcommand{\chiReducedEOFmin}{1.1317}
\newcommand{\chiReducedEOFMCMCmin}{1.1318}
\newcommand{\chiReducedEOFMCMCplus}{0.0014}
\newcommand{\chiReducedEOFMCMCminus}{0.0002}
\newcommand{\chiReducedEOFEXPONENT}{0}
\newcommand{\paramONEEOFmin}{0.008}
\newcommand{\paramONEEOFMCMCmin}{0.008}
\newcommand{\paramONEEOFMCMCplus}{0.035}
\newcommand{\paramONEEOFMCMCminus}{0.032}
\newcommand{\paramONEEOFEXPONENT}{0}
\newcommand{\paramTWOEOFmin}{0.83}
\newcommand{\paramTWOEOFMCMCmin}{0.83}
\newcommand{\paramTWOEOFMCMCplus}{0.02}
\newcommand{\paramTWOEOFMCMCminus}{0.01}
\newcommand{\paramTWOEOFEXPONENT}{0}
\newcommand{\paramTHREEEOFmin}{0.42}
\newcommand{\paramTHREEEOFMCMCmin}{0.42}
\newcommand{\paramTHREEEOFMCMCplus}{0.02}
\newcommand{\paramTHREEEOFMCMCminus}{0.01}
\newcommand{\paramTHREEEOFEXPONENT}{0}
\newcommand{\paramFOUREOFmin}{2.8}
\newcommand{\paramFOUREOFMCMCmin}{2.8}
\newcommand{\paramFOUREOFMCMCplus}{0.1}
\newcommand{\paramFOUREOFMCMCminus}{0.1}
\newcommand{\paramFOUREOFEXPONENT}{0}
\newcommand{\paramFIVEEOFmin}{-2.97}
\newcommand{\paramFIVEEOFMCMCmin}{-2.97}
\newcommand{\paramFIVEEOFMCMCplus}{0.07}
\newcommand{\paramFIVEEOFMCMCminus}{0.06}
\newcommand{\paramFIVEEOFEXPONENT}{0}
\newcommand{\paramSIXEOFmin}{0.95}
\newcommand{\paramSIXEOFMCMCmin}{0.95}
\newcommand{\paramSIXEOFMCMCplus}{0.06}
\newcommand{\paramSIXEOFMCMCminus}{0.03}
\newcommand{\paramSIXEOFEXPONENT}{0}
\newcommand{\paramSEVENEOFmin}{6.766}
\newcommand{\paramSEVENEOFMCMCmin}{6.760}
\newcommand{\paramSEVENEOFMCMCplus}{0.075}
\newcommand{\paramSEVENEOFMCMCminus}{0.006}
\newcommand{\paramSEVENEOFEXPONENT}{0}
\newcommand{\paramEIGHTEOFmin}{0.450}
\newcommand{\paramEIGHTEOFMCMCmin}{0.453}
\newcommand{\paramEIGHTEOFMCMCplus}{0.029}
\newcommand{\paramEIGHTEOFMCMCminus}{0.008}
\newcommand{\paramEIGHTEOFEXPONENT}{0}
\newcommand{\paramNINEEOFmin}{2.968}
\newcommand{\paramNINEEOFMCMCmin}{2.968}
\newcommand{\paramNINEEOFMCMCplus}{0.006}
\newcommand{\paramNINEEOFMCMCminus}{0.006}
\newcommand{\paramNINEEOFEXPONENT}{0}
\newcommand{\paramTENEOFmin}{0.9}
\newcommand{\paramTENEOFMCMCmin}{0.9}
\newcommand{\paramTENEOFMCMCplus}{0.3}
\newcommand{\paramTENEOFMCMCminus}{0.3}
\newcommand{\paramTENEOFEXPONENT}{0}
\newcommand{\paramTENONEEOFmin}{-1.21}
\newcommand{\paramTENONEEOFMCMCmin}{-1.20}
\newcommand{\paramTENONEEOFMCMCplus}{0.03}
\newcommand{\paramTENONEEOFMCMCminus}{0.04}
\newcommand{\paramTENONEEOFEXPONENT}{0}
\newcommand{\paramTENTWOEOFmin}{0.83}
\newcommand{\paramTENTWOEOFMCMCmin}{0.84}
\newcommand{\paramTENTWOEOFMCMCplus}{0.03}
\newcommand{\paramTENTWOEOFMCMCminus}{0.04}
\newcommand{\paramTENTWOEOFEXPONENT}{0}
\newcommand{\paramTENTHREEEOFmin}{74}
\newcommand{\paramTENTHREEEOFMCMCmin}{72}
\newcommand{\paramTENTHREEEOFMCMCplus}{8}
\newcommand{\paramTENTHREEEOFMCMCminus}{2}
\newcommand{\paramTENTHREEEOFEXPONENT}{0}
\newcommand{\paramTENFOUREOFmin}{5.83}
\newcommand{\paramTENFOUREOFMCMCmin}{5.85}
\newcommand{\paramTENFOUREOFMCMCplus}{0.08}
\newcommand{\paramTENFOUREOFMCMCminus}{0.08}
\newcommand{\paramTENFOUREOFEXPONENT}{-10}
\newcommand{\paramTENFIVEEOFmin}{2.2}
\newcommand{\paramTENFIVEEOFMCMCmin}{2.2}
\newcommand{\paramTENFIVEEOFMCMCplus}{0.4}
\newcommand{\paramTENFIVEEOFMCMCminus}{0.3}
\newcommand{\paramTENFIVEEOFEXPONENT}{-10}
\newcommand{\paramTENSIXEOFmin}{113}
\newcommand{\paramTENSIXEOFMCMCmin}{113}
\newcommand{\paramTENSIXEOFMCMCplus}{1}
\newcommand{\paramTENSIXEOFMCMCminus}{1}
\newcommand{\paramTENSIXEOFEXPONENT}{0}
\newcommand{\paramTENSEVENEOFmin}{0.046}
\newcommand{\paramTENSEVENEOFMCMCmin}{0.046}
\newcommand{\paramTENSEVENEOFMCMCplus}{0.001}
\newcommand{\paramTENSEVENEOFMCMCminus}{0.001}
\newcommand{\paramTENSEVENEOFEXPONENT}{0}
\newcommand{\paramTENEIGHTEOFmin}{0.036}
\newcommand{\paramTENEIGHTEOFMCMCmin}{0.037}
\newcommand{\paramTENEIGHTEOFMCMCplus}{0.010}
\newcommand{\paramTENEIGHTEOFMCMCminus}{0.003}
\newcommand{\paramTENEIGHTEOFEXPONENT}{0}
\newcommand{\paramTENNINEEOFmin}{-9999.0}
\newcommand{\paramTENNINEEOFMCMCmin}{-9994.0}
\newcommand{\paramTENNINEEOFMCMCplus}{40.0}
\newcommand{\paramTENNINEEOFMCMCminus}{40.5}
\newcommand{\paramTENNINEEOFEXPONENT}{0}
\newcommand{\paramTENTENEOFmin}{-9998}
\newcommand{\paramTENTENEOFMCMCmin}{-9998}
\newcommand{\paramTENTENEOFMCMCplus}{70}
\newcommand{\paramTENTENEOFMCMCminus}{10}
\newcommand{\paramTENTENEOFEXPONENT}{0}

\newcommand{\chiNSOSmin}{1.0390}
\newcommand{\chiNSOSMCMCmin}{1.0394}
\newcommand{\chiNSOSMCMCplus}{0.0002}
\newcommand{\chiNSOSMCMCminus}{0.0028}
\newcommand{\chiNSOSEXPONENT}{6}
\newcommand{\chiReducedNSOSmin}{1.8451}
\newcommand{\chiReducedNSOSMCMCmin}{1.8458}
\newcommand{\chiReducedNSOSMCMCplus}{0.0004}
\newcommand{\chiReducedNSOSMCMCminus}{0.0049}
\newcommand{\chiReducedNSOSEXPONENT}{0}
\newcommand{\paramONENSOSmin}{2.261}
\newcommand{\paramONENSOSMCMCmin}{2.253}
\newcommand{\paramONENSOSMCMCplus}{0.194}
\newcommand{\paramONENSOSMCMCminus}{0.009}
\newcommand{\paramONENSOSEXPONENT}{0}
\newcommand{\paramTWONSOSmin}{0.455}
\newcommand{\paramTWONSOSMCMCmin}{0.458}
\newcommand{\paramTWONSOSMCMCplus}{0.006}
\newcommand{\paramTWONSOSMCMCminus}{0.014}
\newcommand{\paramTWONSOSEXPONENT}{0}
\newcommand{\paramTHREENSOSmin}{1.015}
\newcommand{\paramTHREENSOSMCMCmin}{1.035}
\newcommand{\paramTHREENSOSMCMCplus}{0.010}
\newcommand{\paramTHREENSOSMCMCminus}{0.183}
\newcommand{\paramTHREENSOSEXPONENT}{0}
\newcommand{\paramFOURNSOSmin}{2.04}
\newcommand{\paramFOURNSOSMCMCmin}{2.05}
\newcommand{\paramFOURNSOSMCMCplus}{0.11}
\newcommand{\paramFOURNSOSMCMCminus}{0.08}
\newcommand{\paramFOURNSOSEXPONENT}{0}
\newcommand{\paramFIVENSOSmin}{-1.84}
\newcommand{\paramFIVENSOSMCMCmin}{-1.84}
\newcommand{\paramFIVENSOSMCMCplus}{0.03}
\newcommand{\paramFIVENSOSMCMCminus}{0.02}
\newcommand{\paramFIVENSOSEXPONENT}{0}
\newcommand{\paramSIXNSOSmin}{1.51}
\newcommand{\paramSIXNSOSMCMCmin}{1.51}
\newcommand{\paramSIXNSOSMCMCplus}{0.06}
\newcommand{\paramSIXNSOSMCMCminus}{0.02}
\newcommand{\paramSIXNSOSEXPONENT}{0}
\newcommand{\paramSEVENNSOSmin}{102}
\newcommand{\paramSEVENNSOSMCMCmin}{102}
\newcommand{\paramSEVENNSOSMCMCplus}{1}
\newcommand{\paramSEVENNSOSMCMCminus}{2}
\newcommand{\paramSEVENNSOSEXPONENT}{0}
\newcommand{\paramEIGHTNSOSmin}{3.96}
\newcommand{\paramEIGHTNSOSMCMCmin}{4.02}
\newcommand{\paramEIGHTNSOSMCMCplus}{0.04}
\newcommand{\paramEIGHTNSOSMCMCminus}{0.28}
\newcommand{\paramEIGHTNSOSEXPONENT}{-10}
\newcommand{\paramNINENSOSmin}{0.018}
\newcommand{\paramNINENSOSMCMCmin}{0.018}
\newcommand{\paramNINENSOSMCMCplus}{0.002}
\newcommand{\paramNINENSOSMCMCminus}{0.002}
\newcommand{\paramNINENSOSEXPONENT}{0}

\newcommand{\chiTSOSmin}{6.6}
\newcommand{\chiTSOSMCMCmin}{6.6}
\newcommand{\chiTSOSMCMCplus}{0.1}
\newcommand{\chiTSOSMCMCminus}{0.2}
\newcommand{\chiTSOSEXPONENT}{5}
\newcommand{\chiReducedTSOSmin}{1.17}
\newcommand{\chiReducedTSOSMCMCmin}{1.17}
\newcommand{\chiReducedTSOSMCMCplus}{0.02}
\newcommand{\chiReducedTSOSMCMCminus}{0.04}
\newcommand{\chiReducedTSOSEXPONENT}{0}
\newcommand{\paramONETSOSmin}{0.12}
\newcommand{\paramONETSOSMCMCmin}{0.12}
\newcommand{\paramONETSOSMCMCplus}{0.03}
\newcommand{\paramONETSOSMCMCminus}{0.04}
\newcommand{\paramONETSOSEXPONENT}{0}
\newcommand{\paramTWOTSOSmin}{0.746}
\newcommand{\paramTWOTSOSMCMCmin}{0.752}
\newcommand{\paramTWOTSOSMCMCplus}{0.061}
\newcommand{\paramTWOTSOSMCMCminus}{0.002}
\newcommand{\paramTWOTSOSEXPONENT}{0}
\newcommand{\paramTHREETSOSmin}{0.48}
\newcommand{\paramTHREETSOSMCMCmin}{0.50}
\newcommand{\paramTHREETSOSMCMCplus}{0.07}
\newcommand{\paramTHREETSOSMCMCminus}{0.18}
\newcommand{\paramTHREETSOSEXPONENT}{0}
\newcommand{\paramFOURTSOSmin}{4.8}
\newcommand{\paramFOURTSOSMCMCmin}{4.7}
\newcommand{\paramFOURTSOSMCMCplus}{0.5}
\newcommand{\paramFOURTSOSMCMCminus}{0.2}
\newcommand{\paramFOURTSOSEXPONENT}{0}
\newcommand{\paramFIVETSOSmin}{-2.8}
\newcommand{\paramFIVETSOSMCMCmin}{-2.8}
\newcommand{\paramFIVETSOSMCMCplus}{0.2}
\newcommand{\paramFIVETSOSMCMCminus}{0.2}
\newcommand{\paramFIVETSOSEXPONENT}{0}
\newcommand{\paramSIXTSOSmin}{1.16}
\newcommand{\paramSIXTSOSMCMCmin}{1.18}
\newcommand{\paramSIXTSOSMCMCplus}{0.04}
\newcommand{\paramSIXTSOSMCMCminus}{0.20}
\newcommand{\paramSIXTSOSEXPONENT}{0}
\newcommand{\paramSEVENTSOSmin}{2.6}
\newcommand{\paramSEVENTSOSMCMCmin}{2.7}
\newcommand{\paramSEVENTSOSMCMCplus}{0.3}
\newcommand{\paramSEVENTSOSMCMCminus}{0.8}
\newcommand{\paramSEVENTSOSEXPONENT}{0}
\newcommand{\paramEIGHTTSOSmin}{1.1}
\newcommand{\paramEIGHTTSOSMCMCmin}{1.0}
\newcommand{\paramEIGHTTSOSMCMCplus}{0.7}
\newcommand{\paramEIGHTTSOSMCMCminus}{0.2}
\newcommand{\paramEIGHTTSOSEXPONENT}{0}
\newcommand{\paramNINETSOSmin}{5.3}
\newcommand{\paramNINETSOSMCMCmin}{5.4}
\newcommand{\paramNINETSOSMCMCplus}{0.3}
\newcommand{\paramNINETSOSMCMCminus}{0.4}
\newcommand{\paramNINETSOSEXPONENT}{0}
\newcommand{\paramTENTSOSmin}{9}
\newcommand{\paramTENTSOSMCMCmin}{9}
\newcommand{\paramTENTSOSMCMCplus}{1}
\newcommand{\paramTENTSOSMCMCminus}{3}
\newcommand{\paramTENTSOSEXPONENT}{0}
\newcommand{\paramTENONETSOSmin}{-0.25}
\newcommand{\paramTENONETSOSMCMCmin}{-0.26}
\newcommand{\paramTENONETSOSMCMCplus}{0.11}
\newcommand{\paramTENONETSOSMCMCminus}{0.04}
\newcommand{\paramTENONETSOSEXPONENT}{0}
\newcommand{\paramTENTWOTSOSmin}{1.59}
\newcommand{\paramTENTWOTSOSMCMCmin}{1.57}
\newcommand{\paramTENTWOTSOSMCMCplus}{0.07}
\newcommand{\paramTENTWOTSOSMCMCminus}{0.06}
\newcommand{\paramTENTWOTSOSEXPONENT}{0}
\newcommand{\paramTENTHREETSOSmin}{63.5}
\newcommand{\paramTENTHREETSOSMCMCmin}{63.1}
\newcommand{\paramTENTHREETSOSMCMCplus}{11.1}
\newcommand{\paramTENTHREETSOSMCMCminus}{0.7}
\newcommand{\paramTENTHREETSOSEXPONENT}{0}
\newcommand{\paramTENFOURTSOSmin}{5.63}
\newcommand{\paramTENFOURTSOSMCMCmin}{5.66}
\newcommand{\paramTENFOURTSOSMCMCplus}{0.08}
\newcommand{\paramTENFOURTSOSMCMCminus}{0.25}
\newcommand{\paramTENFOURTSOSEXPONENT}{-10}
\newcommand{\paramTENFIVETSOSmin}{2.41}
\newcommand{\paramTENFIVETSOSMCMCmin}{2.40}
\newcommand{\paramTENFIVETSOSMCMCplus}{0.01}
\newcommand{\paramTENFIVETSOSMCMCminus}{0.17}
\newcommand{\paramTENFIVETSOSEXPONENT}{-10}
\newcommand{\paramTENSIXTSOSmin}{112}
\newcommand{\paramTENSIXTSOSMCMCmin}{112}
\newcommand{\paramTENSIXTSOSMCMCplus}{2}
\newcommand{\paramTENSIXTSOSMCMCminus}{2}
\newcommand{\paramTENSIXTSOSEXPONENT}{0}
\newcommand{\paramTENSEVENTSOSmin}{0.0429}
\newcommand{\paramTENSEVENTSOSMCMCmin}{0.0435}
\newcommand{\paramTENSEVENTSOSMCMCplus}{0.0043}
\newcommand{\paramTENSEVENTSOSMCMCminus}{0.0009}
\newcommand{\paramTENSEVENTSOSEXPONENT}{0}
\newcommand{\paramTENEIGHTTSOSmin}{0.056}
\newcommand{\paramTENEIGHTTSOSMCMCmin}{0.057}
\newcommand{\paramTENEIGHTTSOSMCMCplus}{0.004}
\newcommand{\paramTENEIGHTTSOSMCMCminus}{0.008}
\newcommand{\paramTENEIGHTTSOSEXPONENT}{0}
\newcommand{\paramTENNINETSOSmin}{0.637}
\newcommand{\paramTENNINETSOSMCMCmin}{0.635}
\newcommand{\paramTENNINETSOSMCMCplus}{0.006}
\newcommand{\paramTENNINETSOSMCMCminus}{0.004}
\newcommand{\paramTENNINETSOSEXPONENT}{0}
\newcommand{\paramTENTENTSOSmin}{0.849}
\newcommand{\paramTENTENTSOSMCMCmin}{0.848}
\newcommand{\paramTENTENTSOSMCMCplus}{0.005}
\newcommand{\paramTENTENTSOSMCMCminus}{0.007}
\newcommand{\paramTENTENTSOSEXPONENT}{0}

\shorttitle{Dust Distribution Around $\beta$ Pictoris} 
\shortauthors{Ahmic et al.}

\begin{document}

\title{Dust Distribution in the $\beta$ Pictoris Circumstellar Disks}

\author{Mirza Ahmic\footnote{e-mail: ahmicm@math.mcmaster.ca}, Bryce Croll, Pawel Artymowicz\footnote{University of Toronto at Scarborough, Toronto, ON M1C 1A4, Canada}}
\affil{Department of Astronomy \& Astrophysics, University of Toronto,
    Toronto, ON M5S 3H8, Canada}

\begin{abstract}

We present 3-D  models of dust distribution around $\beta$ Pictoris that produce the 
best fits to the Hubble Space Telescope Advanced Camera for Surveys' (HST/ACS) images obtained by Golimowski and co-workers.
We allow for the presence of either one or two
separate axisymmetric dust disks. The density models are analytical, radial two-power-laws joined smoothly at a cross-over radius with density exponentially decreasing away from the mid-plane 
of the disks. Two-disk models match the data best, yielding a reduced $\chi^2$ of $\sim$1.2. 
Our two-disk model reproduces many of the asymmetries reported in the literature and 
suggests that it is the secondary (tilted) disk which is largely responsible for them.
Our model suggests that the secondary disk is not constrained to the inner regions of the system 
(extending out to at least 250\,AU) and that it has a slightly larger total area of dust than the primary, as a result of slower fall-off of density with radius
and height. This surprising result
raises many questions about the origin and dynamics of such a pair of disks. The disks 
overlap, but can coexist owing to their low optical depths and therefore long mean collision times.
We find that the two disks have dust replenishment
times on the order of $10^4$ yr at $\sim$100 AU, hinting
at the presence of planetesimals that are responsible for the production of $2^{\mathrm{nd}}$ generation dust. A plausible conjecture, which needs to be confirmed by physical modeling of the collisional dynamics of bodies in the disks, is that the two observed disks are derived from underlying planetesimal disks; such disks would be anchored by the gravitational influence of planets located at less than 70 AU from $\beta$ Pic that are themselves in slightly inclined orbits.    

\end{abstract}


\section{Introduction}

Owing to its proximity ($19.28\pm0.19$pc; \citealt{cri97}), $\beta$ Pictoris is one of the best studied examples of a Main Sequence (MS) 
star with a circumstellar dust disk.
$\beta$ Pic was shown to have a substantial mid- and far-infrared excess through observations using 
the Infrared Astronomical Satelite (IRAS) 
(\citealt{Aum84}; \citealt{Aum85}; \citealt{gil86}), believed to be 
indicative of thermal radiation from $\sim$100 K dust orbiting the star.
Follow-up imaging by \citet{smi84} revealed an almost edge-on system with
circumstellar nebulosity due to light scattered from the dust around the star.
Since then, many investigators have mapped the surface brightness and color of the disk in optical wavelenghts \citep{smi87, art89, les93, gol93, kal95, hea00, gol06}. 

The extensive studies of the disk have also revealed several asymmetries in the scattered light contours produced by the dust surrounding $\beta$ Pic. These were summarized by \citet{kal95} and include:
1. the size asymmetry - the north-east (NE) extension of the disk stretches further out from $\beta$ Pictoris than the south-west (SW) extension,
2. the surface brightness asymmetry - the brightness profile along the disk's spine (midplane) is broader for the NE than the SW extension,
3. the width asymmetry - the SW extension is thicker than the NE extension,
4. the wing tilt asymmetry - the two extensions of the disk are not perfectly aligned (i.e. the 
difference in the position angles of the midplanes of the opposing extensions is less than $180\degr$)
and 5. the butterfly asymmetry - isophote curvature is asymmetric across the midplane of the disk, with the asymmetry itself inverted across the minor-projection axis. Several attempts have been made to explain why these asymmetries arise. \citet{kal95} noted that the size and surface brightness asymmetries might be a direct result of the width asymmetry.
Since the SW extension of the disk is more vertically extended, the SW midplane might appear fainter then the NE midplane and thus explain why the latter appears to have a larger radial size. 

Further insight into the $\beta$ Pic system has come from optical and mid-infrared observations of its disk.
\citet{hea00} and \citet{gol06} imaged the disk with high-resolution HST 
optical cameras and detected a warp in the inner part of the disk ($\sim$20--100 AU from $\beta$ Pic) indicating the presence of a secondary disk inclined at $\sim 5\degr$ to the primary \citep{hea00, gol06}. Intriguingly, \citet{gol06} noted that the projected spine of the secondary is aligned with the isophotal inflections (the butterfly asymmetry) reported at large distances from the star. \citet{gol06} also placed constraints on the brightness profiles along the spines of the primary, secondary and composite (combined primary and secondary) disks. They found that the primary disk has a steeper midplane profile than the secondary at greater distances from the star and that the composite profile is best described with four power laws, in contrast
to previous modelling efforts that employed two power laws (eg. \citealt{kal95}). To probe the regions of $\beta$ Pic's disk inaccessible with optical instruments, \citet{wah03} imaged the disk in
the mid-infrared ($17.9\, \micron$). They reported a set of rings embedded in $\beta$ Pic's disk whose radii vary from 14 to 82\,AU and whose configuration suggests the presence of multiple planets. However, the ring structure is
uncertain, since \citet{gol06} failed to observe them in scattered  light. 

Strong motivation for the study of dust distribution in disks is spurred by the fact that such a study reveals important clues about the presence of unseen larger bodies (planetesimals and planets). 
For example, it was suggested that the HST warp \citep{hea00} is 
caused by the presence of a giant planet on a slightly inclined orbit less than 50\,AU from the primary (ex. \citealt{mou97}).
This prediction was possibly vindicated recently by the announcement of an $8\mathrm{M}_\mathrm{Jup}$ candidate companion
separated by 8\,AU from $\beta$ Pic, which, if confirmed, would make it one of the first planets imaged orbiting an A type MS star, and the
planet imaged with the smallest projected physical separation to-date \citep{lag08}.

The system's many peculiarities have yet to be adequately
explained, and thus continued studies of the dust distribution are a necessity that
might lead to the discovery of additional companions. Given the possibility of dust migration \citep{tak01}  and/or dust avalanches \citep{gri07}, 
the dust and the planetesimal/planet distributions may not be the same.
However, detailed modeling of micron-sized dust including its dynamical interaction with radiation and dust-dust collisions, 
is a prerequisite for understanding the distribution of larger bodies. Our work provides the first step for such prospective modeling, the description of the dust distribution. 

In this paper we present the best-fit two-disk model to the images
obtained by \citet{gol06}. In section 2 we discuss the model assumptions as well as the ACS images that are used for the fitting procedure.
We dedicate section 3 to the results and section 4 to the discussion and analysis.  

\section{Modeling and Data}

\subsection{Multi-Parametric Model}
 
We use an axisymmetric multi-parameter model to describe the distribution of the dust surrounding $\beta$ Pic.
Its simplicity and the fact
that it has been successfully applied in the past on more than one occasion
(eg. \citealt{art89, kal95}) allow one to quickly proceed from the numerical development stage to testing the model on astronomical data. The improvements on the previous applications of the multi-parametric model include the introduction of an additional axis of rotation (which allows us to model disks whose axis can point in any direction), implementation of a $\chi^2$ minimization algorithm (in this case a Markov chain Monte Carlo method) and the addition of the secondary disk which we fit simultaneously with the primary. Furthermore, recent \citet{gol06} observations have produced the most photometrically accurate maps to date which allow one to probe dust distribution in all directions (specifically at large scale heights), allowing one to introduce additional constraints on the model.

In order to describe the dust distribution, we follow \citet{art89}. We assume that both disks are axisymmetric. In cylindrical co-ordinates ($r$,$z$,$\theta$) then, the scattering cross section per unit volume is given by:

\begin{equation}
	\frac{d\tau}{ds} = \frac{\tau(r)}{W(r)}e^{-\left(| z|/W(r)\right)^p}
\end{equation}

This form arises as a result of the vertical exponential drop off of the dust density.
The rate of this drop off is determined by the parameter $p$. $W(r)$ is the width profile:

\begin{equation}
	W(r)= A_0\,R_m\left(\frac{r}{R_m}\right)^{\gamma}
\end{equation}

This profile guarantees that the disk is thicker at greater radii than at smaller radii. If $\gamma$, the flare index, is more than 1, the disk will flare. The scale height $A_0$ ensures that the disk has a specific thickness to radius ratio at $r={R_m}$, the location of the power law break describing the vertical optical thickness. 

The vertical optical thickness is proportional to the function $\tau(r)$. $\tau(r)$ has the following dependance on distance from the centre of the star:
\begin{equation}
 \tau(r) = \frac{ \eta }{\sqrt{ (r/R_m)^{-2\alpha} + (r/R_m)^{-2\beta} }}
\end{equation}

The inner radial index $\alpha$ and the outer radial index $\beta$ suppress the vertical optical thickness below and above the characteristic radius $R_m$. Only two indices are used since the midplane surface brightness profile of the primary disk is known to have a single break and thus requires two power laws to describe it (\citealt{art89, kal95, gol06}).

To compare the brightness contours produced by our model with images of $\beta$ Pic's disk, we convert the cylindrical symmetric dust distribution into the observed brightness isophotes in the following manner. Light intensity along the line of sight through the disk is given by:

\begin{equation}
	I(x,y)= \int{\frac{\frac{d\tau}{ds}(x,y,l)\,f(\theta)\,dl}{4\pi(r^2+z^2)}} 
\end{equation}

Here $x$ and $y$ are the pixel co-ordinates of an image and l is the depth along the line of sight. In the above equation, flux entering a unit volume having co-ordinates ($r$,$z$) is $L_\nu
/[4\pi(r^2+z^2)]$, where $L_\nu$ denotes the luminosity of $\beta$ Pic at frequency $\nu$. This luminosity, along with the distance to the $\beta$ Pic, is hidden inside our
normalization parameter $\eta$. $f(\theta)$ represents the phase function and is described in section 2.2. The formulae for $r$, $z$ and $\theta$ as functions of $x$, $y$ and $l$ follow from the appropriate co-ordinate transformations.
     
\begin{equation}
	r=\left[(l\,\cos i+y\,\sin i)^2+x^2\right]^{1/2}
\end{equation}
\begin{equation}
	z=y\,\cos i \,-l\,\sin i 
\end{equation}
\begin{equation}
	\theta=\mathrm{cos}^{-1}\left(-l/(x^2+y^2+l^2)^{1/2}\right)
\end{equation} 

The theoretical image obtained via the described transformation will show a disk that is aligned with the major projection axis. Angle $\phi$ is introduced so that this image can be rotated around the line-of-sight until it is aligned with a disk, which may not lie along the major projection axis (which is certainly the case for the secondary).
  
\subsection{Phase Functions}

In order to obtain a theoretical image, one must also include a phase function [$f(\theta)$ in equation 4] which describes
how the light is scattered in different directions compared to the line-of-sight by the orbiting dust\footnote{We assume the same optical properties of dust everywhere.}. For our simulations we used two scattering profiles, the empirical zodiacal \citep{lei76} and the Henyey-Greenstein \citep{hen41}. The former profile describes light scattered by the midsized particles (10-100\,$\micron$) with rough surfaces \citep{sci82} and it strongly favors forward scattering. The empirical zodiacal is proportional to:
\begin{equation}
	f(\theta)=0.3(0.2+\theta/2)^{-3}+1.4(\theta/3.3)^4+0.2\,.
\end{equation} 

\begin{figure}[tbp]
\begin{center}
\includegraphics[angle=270,scale=0.7]{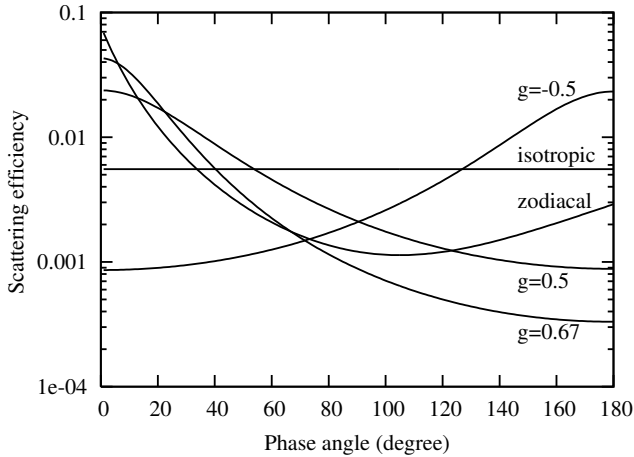}
\end{center}
\caption{The scattering efficiency as a function of phase angle is plotted for empirical zodiacal \citep{lei76} and several Henyey-Greenstein fits \citep{hen41}, including isotropic scattering (g=0).}\label{SCAT}
\end{figure}

The Henyey-Greenstein scattering profile is an analytical phase function commonly used to describe light scattered by small grains: 
\begin{equation}
	f(\theta)=\frac{1-g^2}{4\pi(1+g^2-2g \cos\theta)^{3/2}},
\end{equation}
where $-1<g<1$ is the profile asymmetry parameter. Positive values of $g$ produce forward scattering, negative backward while $g=0$ gives us
the isotropic scattering. In Fig.\ref{SCAT} we present the zodiacal light and several Henyey-Greenstein phase functions.
In the simulations (as well as Fig.\ref{SCAT}) the phase functions are normalized so that their intergrals
over the full solid angle are equal. This is done so that a meaningful comparison can be made between $\eta$ parameters obtained from different simulations.

\subsection{Data}

In order to constrain the two-disk model of $\beta$ Pic's dust distribution, we use the \citet{gol06} ACS Hubble Space Telescope (HST) images. The ACS' excellent optics and HST's well studied PSF allow one to produce images whose quality (photometric accuracy and morphological details) are unparalleled by any other hardware combination currently in operation. Furthermore, the ACS' wide field-of-view (FOV) [$\sim29\arcsec\times26\arcsec$], 
its ability to access large scale heights [$\sim1.5\arcsec$ circular coronographic mask only blocks out the innermost 30\,AU] and
its ability to use three filters [F435W(B), F606W(broad V) and F814(broad I)] to probe chromatic dependence introduce additional constrains on our model that would have been impossible with previous observations.

To discriminate between the features in the image associated with the disk and those intrinsic to the coronographic PSF of the disk and $\beta$ Pic itself, \citet{gol06} observed the disk at two roll angles and obtained images of a star with similar colors ($\alpha$ Pic) prior to every exposure. Following the appropriate subtractions, the combined images were deconvolved with synthetic PSFs produced by the Tiny Tim software package\footnote{For a more detailed discussion of the reduction procedure, we refer the reader to \citet{gol06}}. Throughout the reduction process \citet{gol06} keep track of errors (both random and systematic) associated with each pixel.
This allows them to produce meaningful error maps that include both the systematic and statistical uncertainties in their data; we use 
these error maps for our $\chi^2$ minimization. The number of data points in these images once the coronograph (inner 30\,AU) is masked out is 563136.

\subsection{Markov chain Monte Carlo}

We employ Markov chain Monte Carlo (MCMC; \citealt{Ford}) fitting as described for our purposes in \citealt{CrollMCMC}.
MCMC fitting is a computationally efficient method to perform a full Bayesian analysis of complicted problems where one is
required to fit a great number of often correlated parameters. MCMC fitting effectively samples the posterior parameter distribution
by performing an n-step intelligent random walk around the $K$-dimensional paramater space of interest, while recording the $\chi^{2}$ at each
point in the intelligent random walk.
In addition, MCMC fitting allows one to explore correlations in one's
fitting-parameters - the end result
is that one is able to return realistic best-fit and error estimates even if the 
parameters of interest are correlated.
MCMC fitting is thus well-suited to fit the \citet{gol06} observations of the $\beta$ Pic disk with our model
as it features a large number of parameters ($K$=18-20 parameters for the two-disk case)
a select number of which display modest correlations. 

Flat priors in all parameters are used. We run our MCMC chains until the \citet{Gelman} statistic is close to unity for all parameters.
To determine our best-fit and 1-$\sigma$ uncertainties, we use the marginalized likelihood method as described in \citet{CrollMCMC}.

\subsubsection{Accounting for the large number of data points}
\label{SecLargeNumberOfDataPoints}

Given the large number of data-points ($>$500000) in the ACS images of $\beta$ Pic and the impressive accuracy of this data minute changes in one's
model will result in a statistically significant increase in $\chi^{2}$. 
These small increases in $\chi^{2}$ indicate a statistically significant poorer fit in the case that one's model is an accurate representation of the 
physical reality of the disk in this case. However, in our case it should be noted 
that even in the best-case scenario our model is likely only an
approximation of the actual physical makeup of the $\beta$ Pic disk; 
even our best-case scenario cannot account for the numerous asymmetries in the disk (e.g. Figure \ref{TOTRES9AND18}).
Furthermore, although the error estimates provided by \citet{gol06} attempt to include both the systematic and statistical
uncertanty in the ACS data, they could fail to account for correlations in the data that could add an extra systematic component.
For these reasons we feel that a simplistic $\chi^2$ could 
underestimate the true uncertainties in our fitted parameters. For these reasons 
we have scaled up the uncertainties in our fitted parameters by a factor of 10, as we believe these give a more accurate indication
of the true uncertainty in our data.

\section{Results}

In the following sections we discuss optimal parameters for a single and two-disk dust distributions as well as several scattering phase functions. Since both the dust distribution and the phase function are responsible for the scattered light profile, both components are investigated here 
and modelled with our simulations. However, in order to minimize the number of free parameters, we first explore the parameter space using
a zodiacal light phase function for scattering. In section 3.3 we expand on our work by exploring other phase distributions. 

\subsection{One Disk vs. Two Disk Model}

To test both the fitting procedure and the goodness-of-fit of our two-disk fits, we first constrained the 
parameters describing the dust distribution via a single disk and a zodiacal light scattering profile. The zodiacal light profile has been used in the past (eg. \citealt{kal95}) and was chosen in order to 
minimize the number of free parameters.  
Table \ref{SINGPARCOMP} lists these results as well as those obtained in the past using similar axisymmetric models. We do not note any significant disagreements and suggest that the minor differences between our and past results might be due to several factors including: 1. that previous fitting efforts were performed with simpler models (eg. \citealt{kal95}), 2. that the previous data used for modelling efforts was of poorer quality, 3. that various coronographic implementations employed in previous observations did not allow the observer to reliably probe large scale heights, 4. the usage of different phase functions and lastly 5. that past modellers applied trial and error methods to explore the parameter space instead of more sophisticated techniques. 

\begin{deluxetable*}{lccc}[tbp]
\tabletypesize{\footnotesize}
\tablecaption{Single disk parameter comparison}
\tablewidth{0pt}
\tablehead{
\colhead{Parameter} & \colhead{\citet{kal95}\tablenotemark{a,c}} & \colhead{Artymowicz (1998)\tablenotemark{b,c}} & \colhead{This paper}
}
\startdata
Inclination & $2\degr<i<5\degr$ & $i=1.3\degr$ & $i=2.3\degr$ \\
Vertical distribution & $0.7<p<2.0$ & $p=0.7$ & $p=0.5$ \\
Radius of the power law break & $R_m=100\,\mathrm{AU}$ & $R_m=120\,\mathrm{AU}$ & $R_m=102\,\mathrm{AU}$ \\
Inner radial index & N/A & $\alpha=2.0$ & $\alpha=2.0$ \\
Outer radial index & $-3.4<\beta<-2.8$ & $\beta=-3.0$ & $\beta=-1.8$ \\
Flare index & $1.1<\gamma<1.6$ & $\gamma=0.75$ & $\gamma=1.5$ \\
Scale height & $0.05<A_0<0.10$ & $A_0=0.055$ & $A_0=0.02$ \\
\enddata
\tablenotetext{a}{Parameter values converted from \citet{kal95} to match our model values.}
\tablenotetext{b}{Personal communique. Results obtained by fitting \citet{hea00} HST STIS images.}
\tablenotetext{c}{Fits obtained by trial and error.}
\label{SINGPARCOMP}
\end{deluxetable*}

\begin{figure*}[tbp]
\begin{center}
\includegraphics[angle=270,scale=0.3]{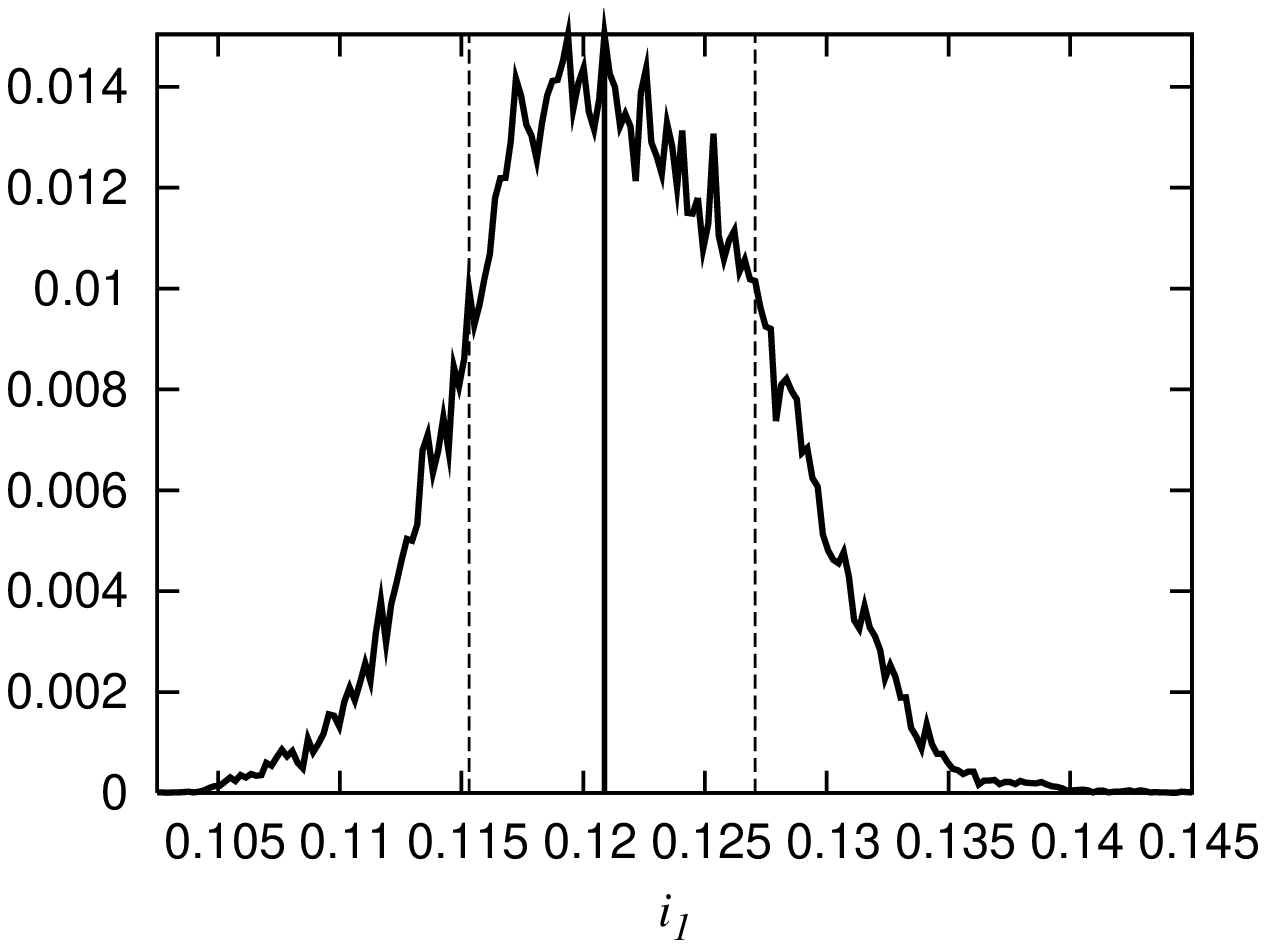}
\includegraphics[angle=270,scale=0.3]{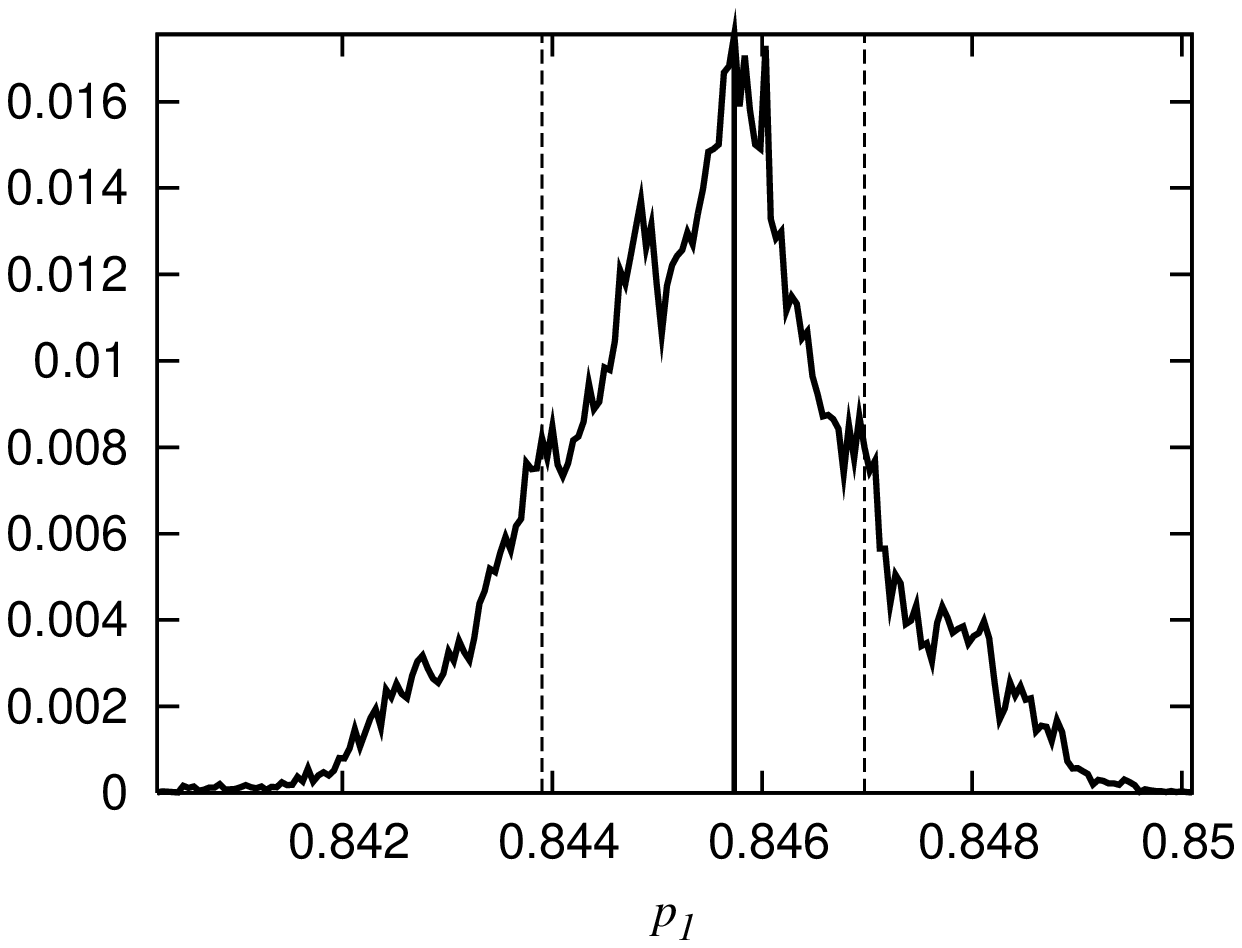}
\includegraphics[angle=270,scale=0.3]{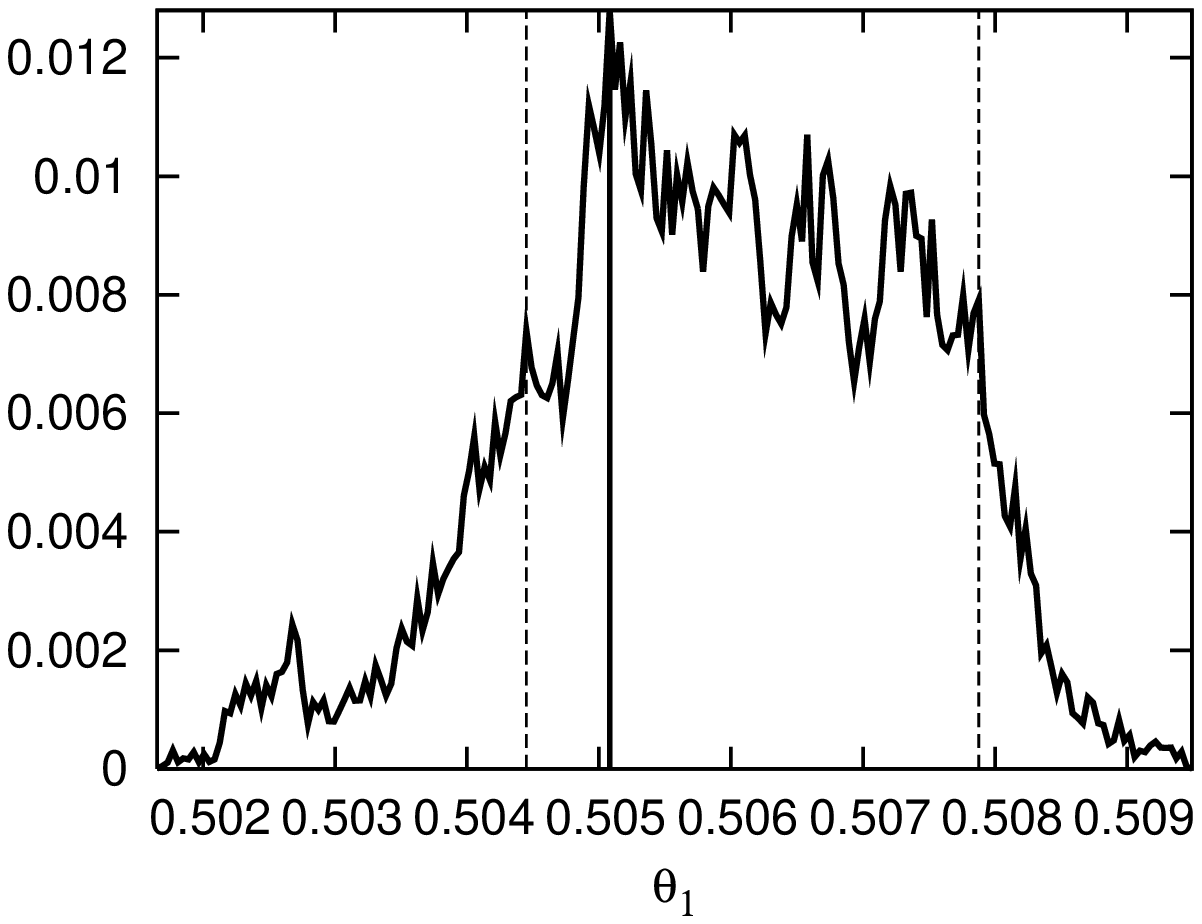}
\includegraphics[angle=270,scale=0.3]{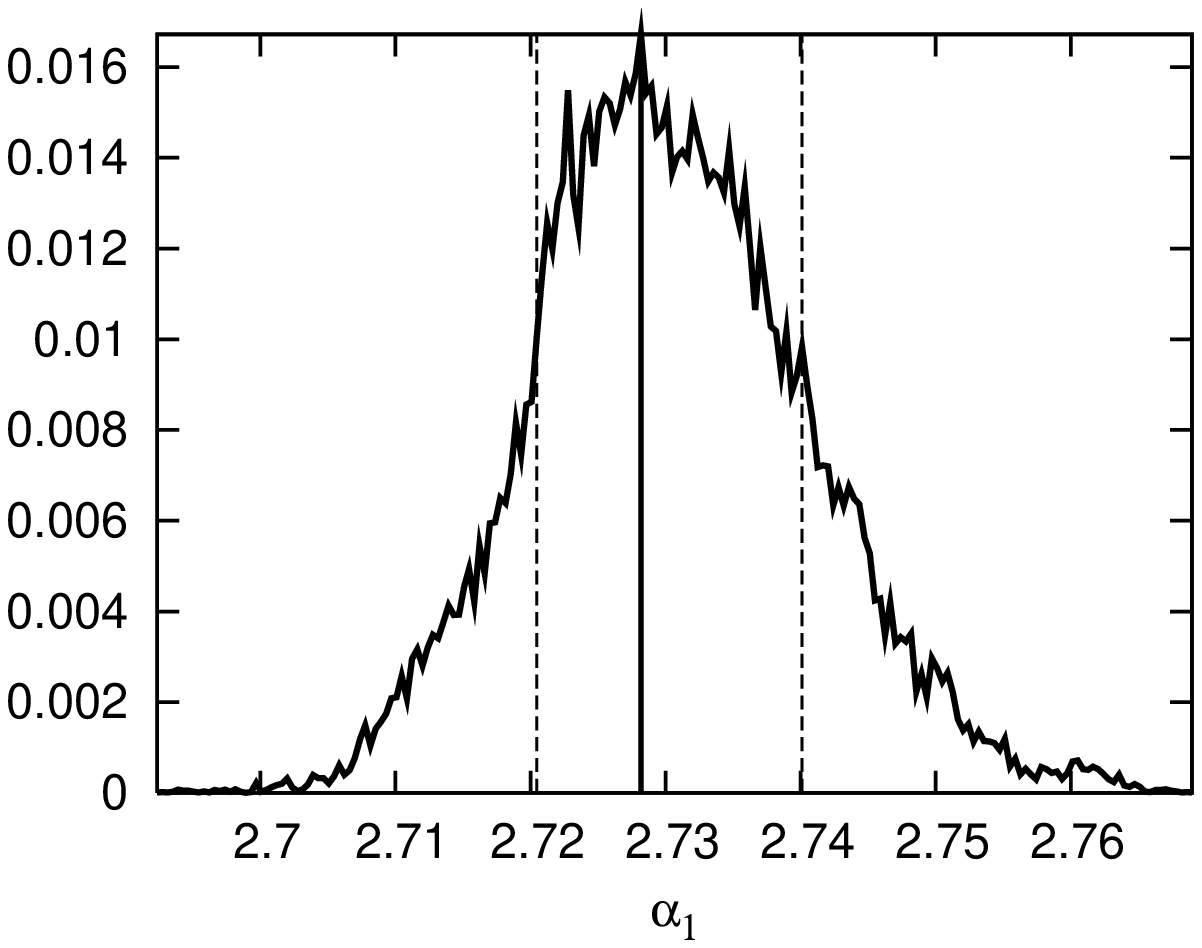}
\includegraphics[angle=270,scale=0.3]{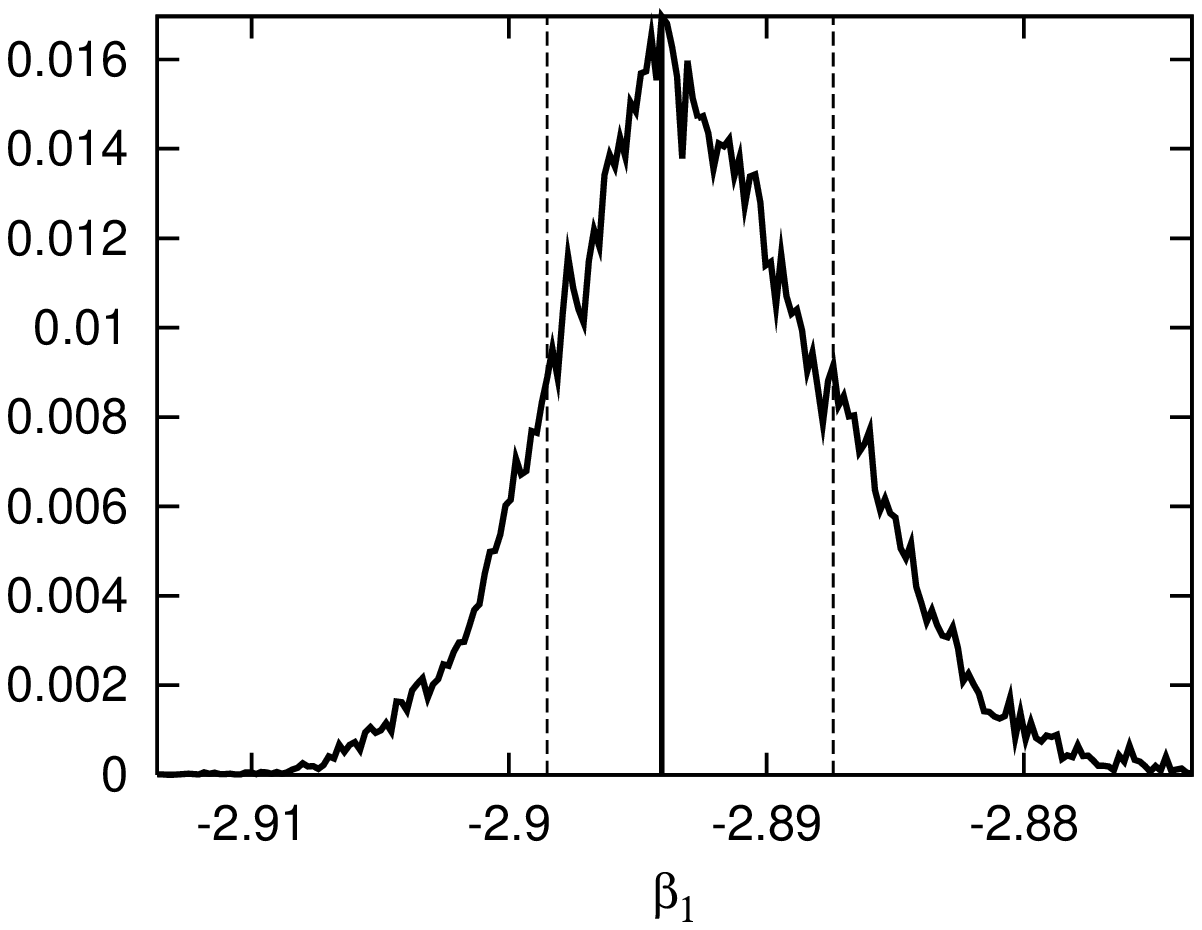}
\includegraphics[angle=270,scale=0.3]{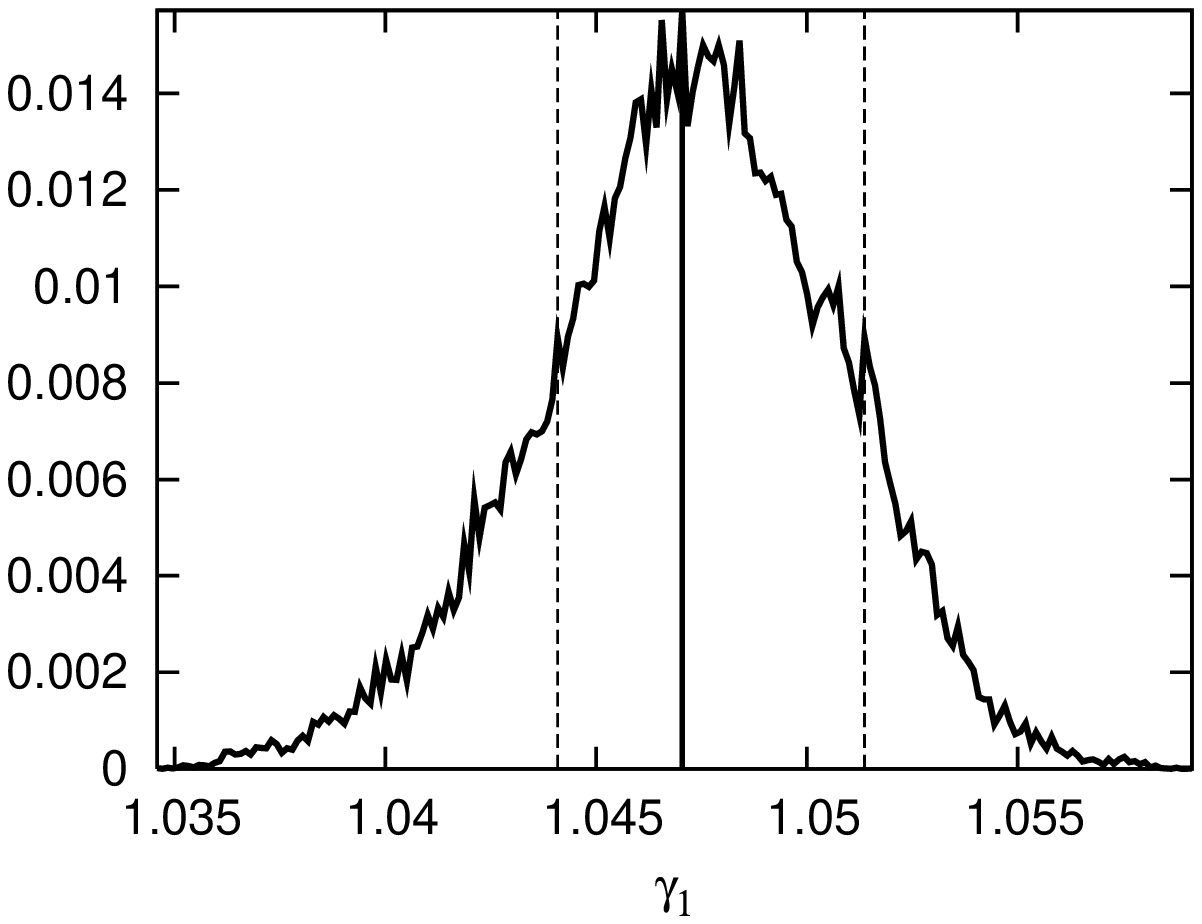}
\includegraphics[angle=270,scale=0.3]{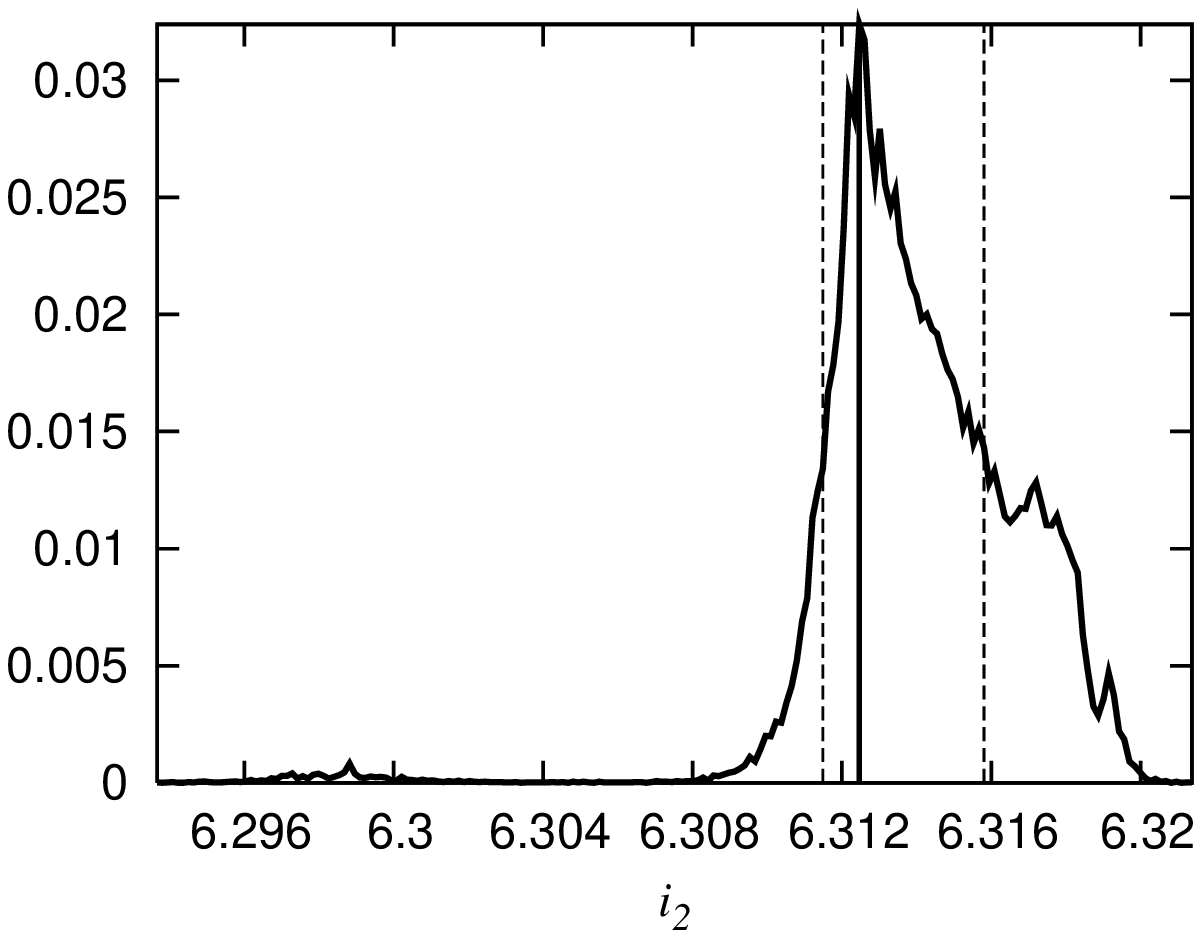}
\includegraphics[angle=270,scale=0.3]{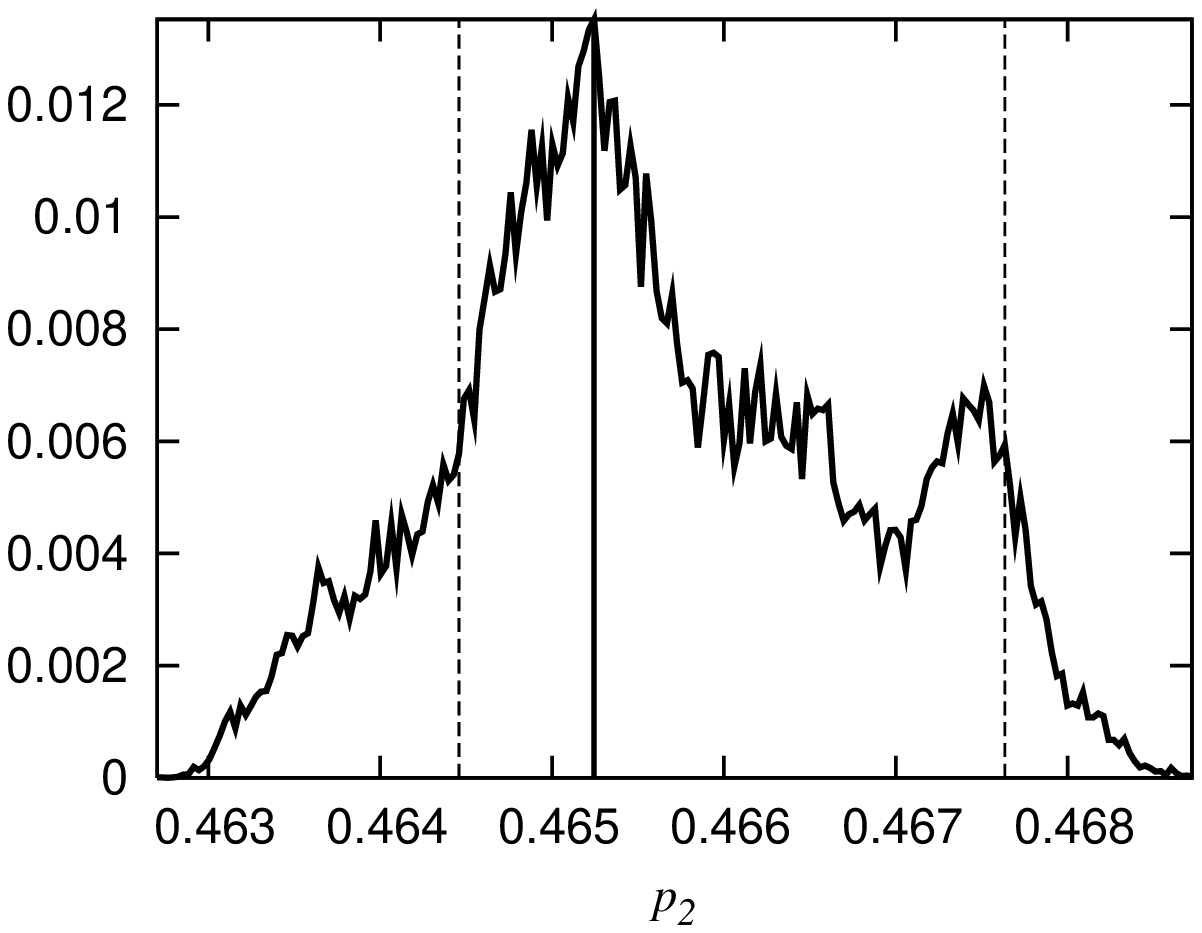}
\includegraphics[angle=270,scale=0.3]{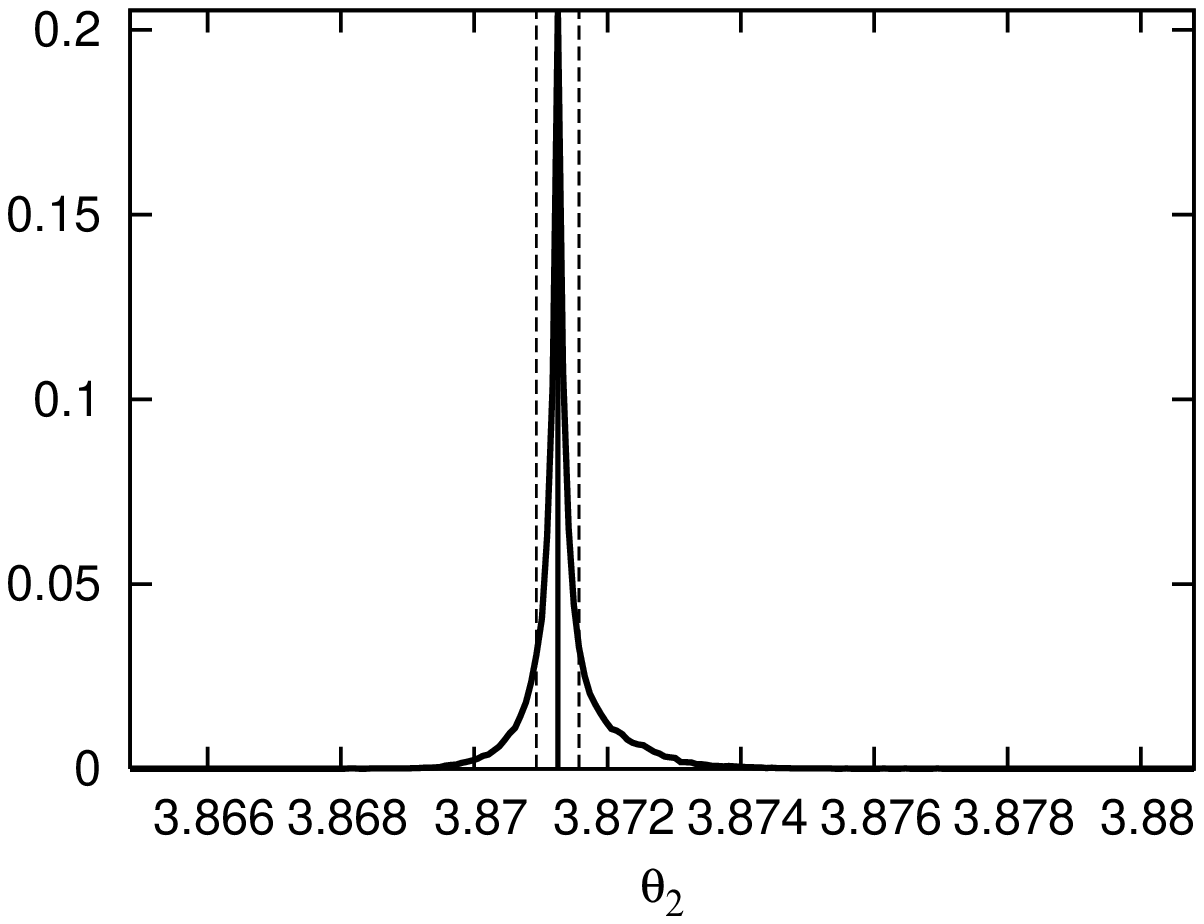}
\includegraphics[angle=270,scale=0.3]{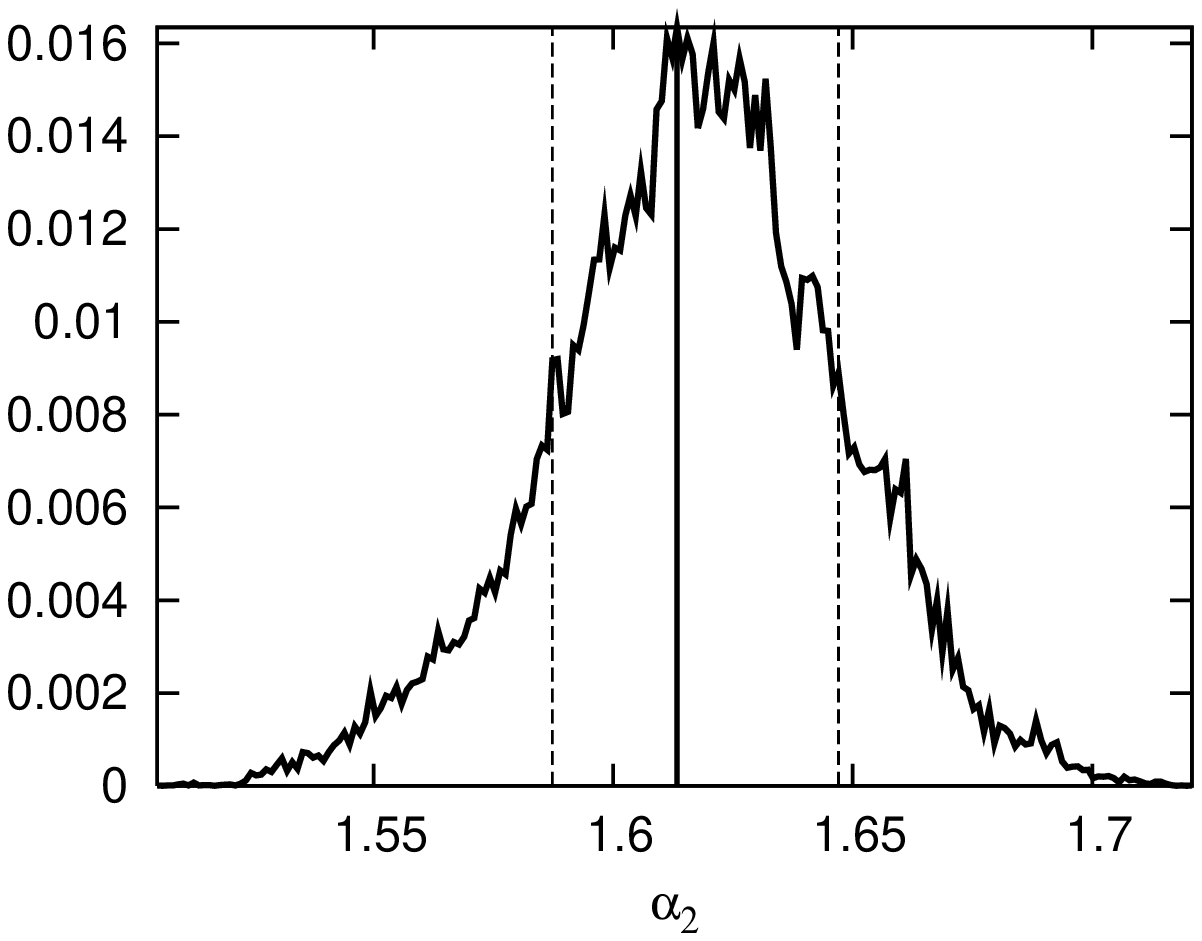}
\includegraphics[angle=270,scale=0.3]{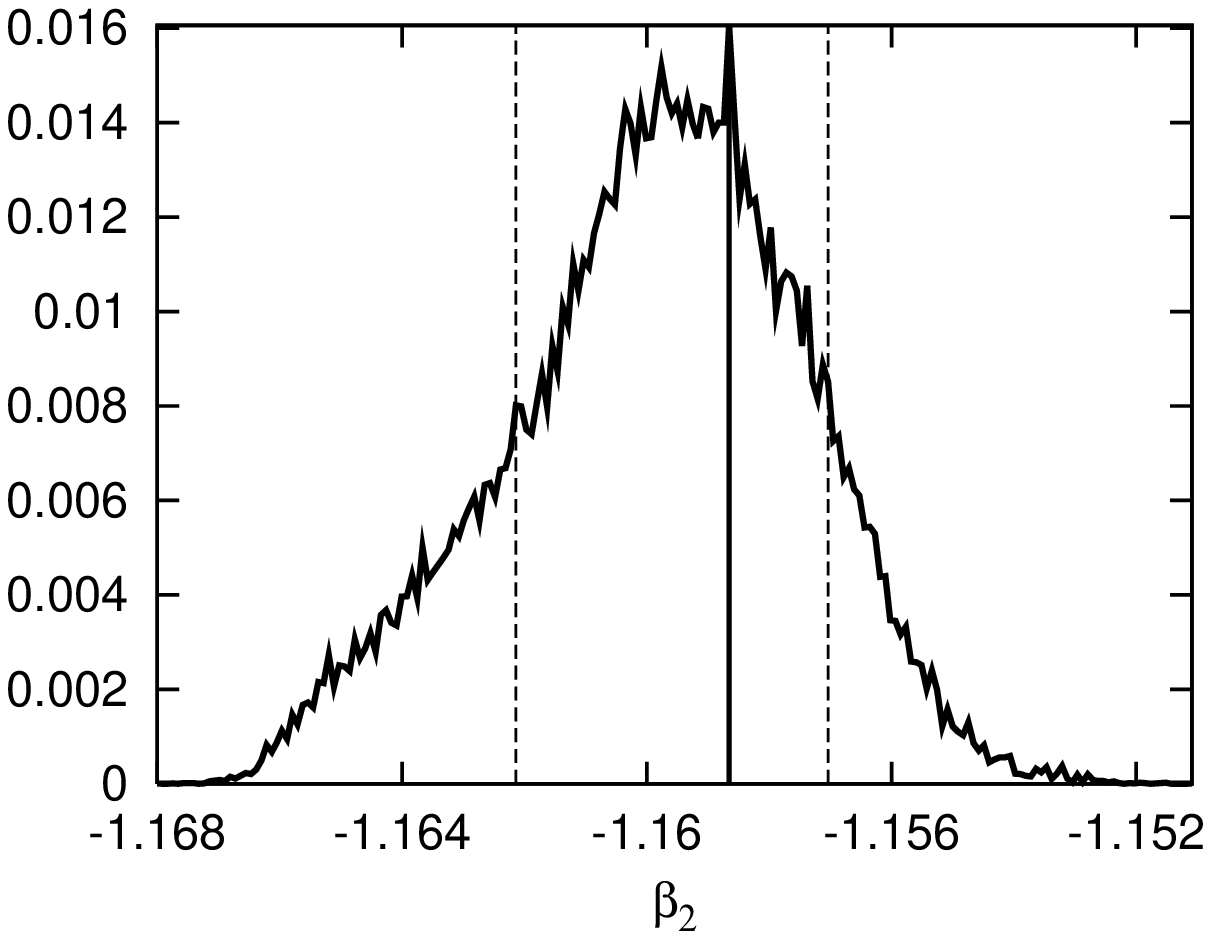}
\includegraphics[angle=270,scale=0.3]{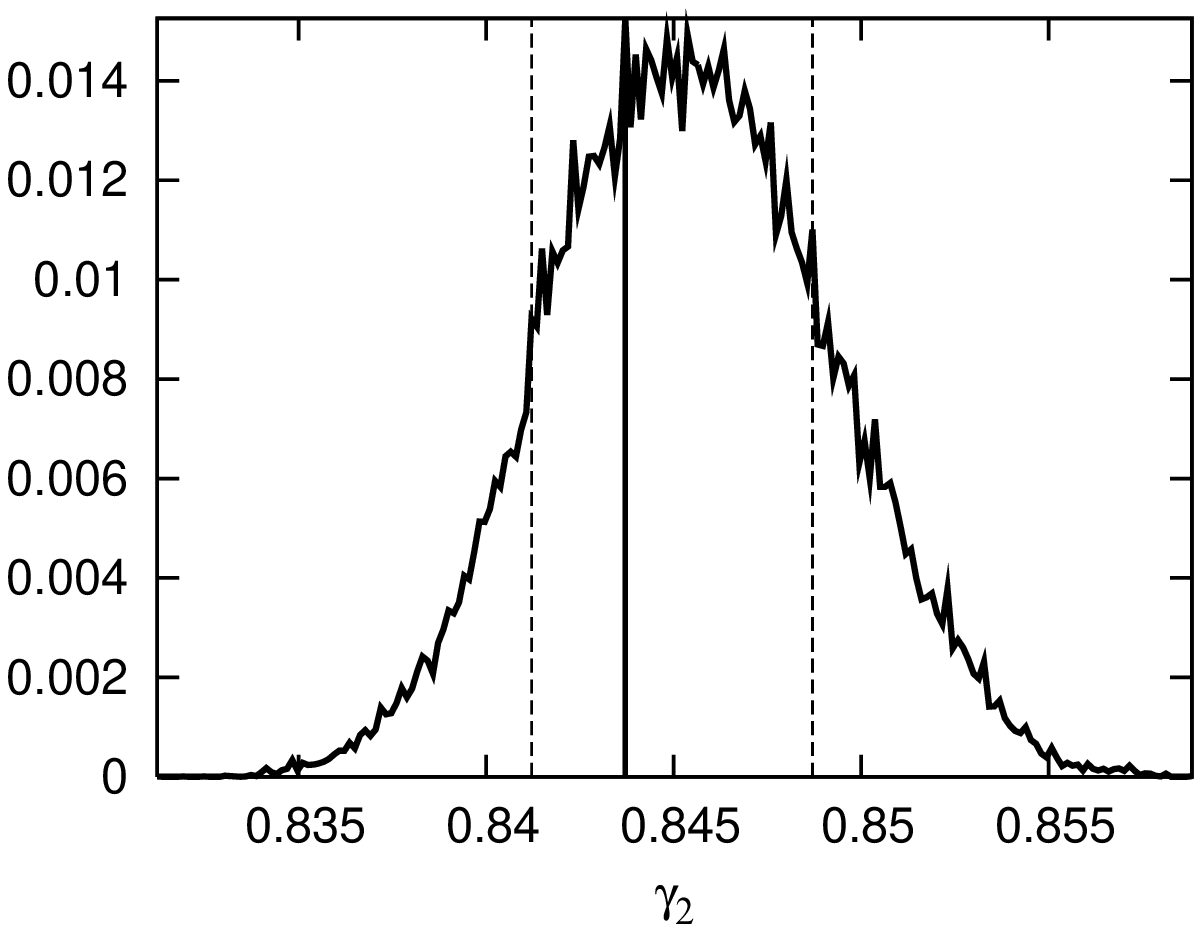}
\includegraphics[angle=270,scale=0.3]{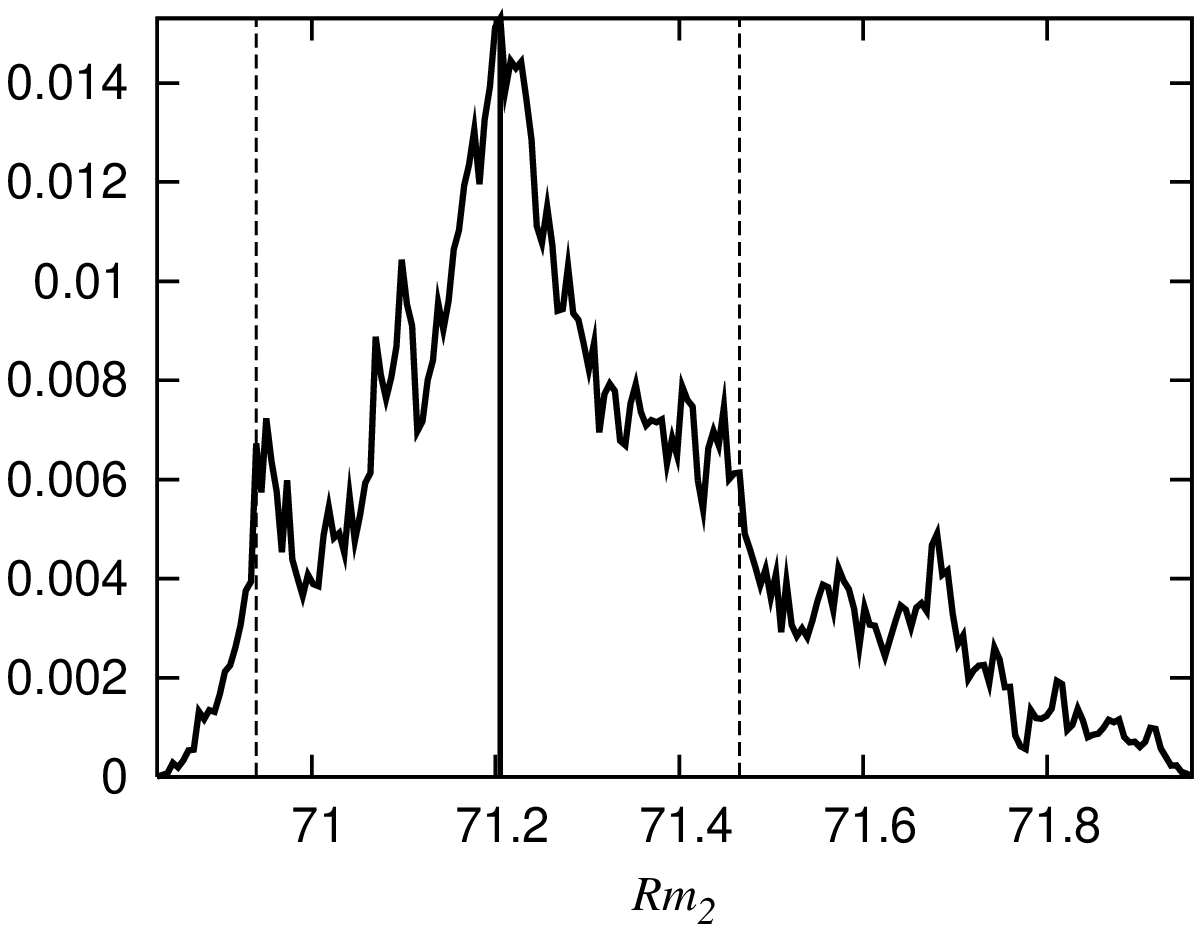}
\includegraphics[angle=270,scale=0.3]{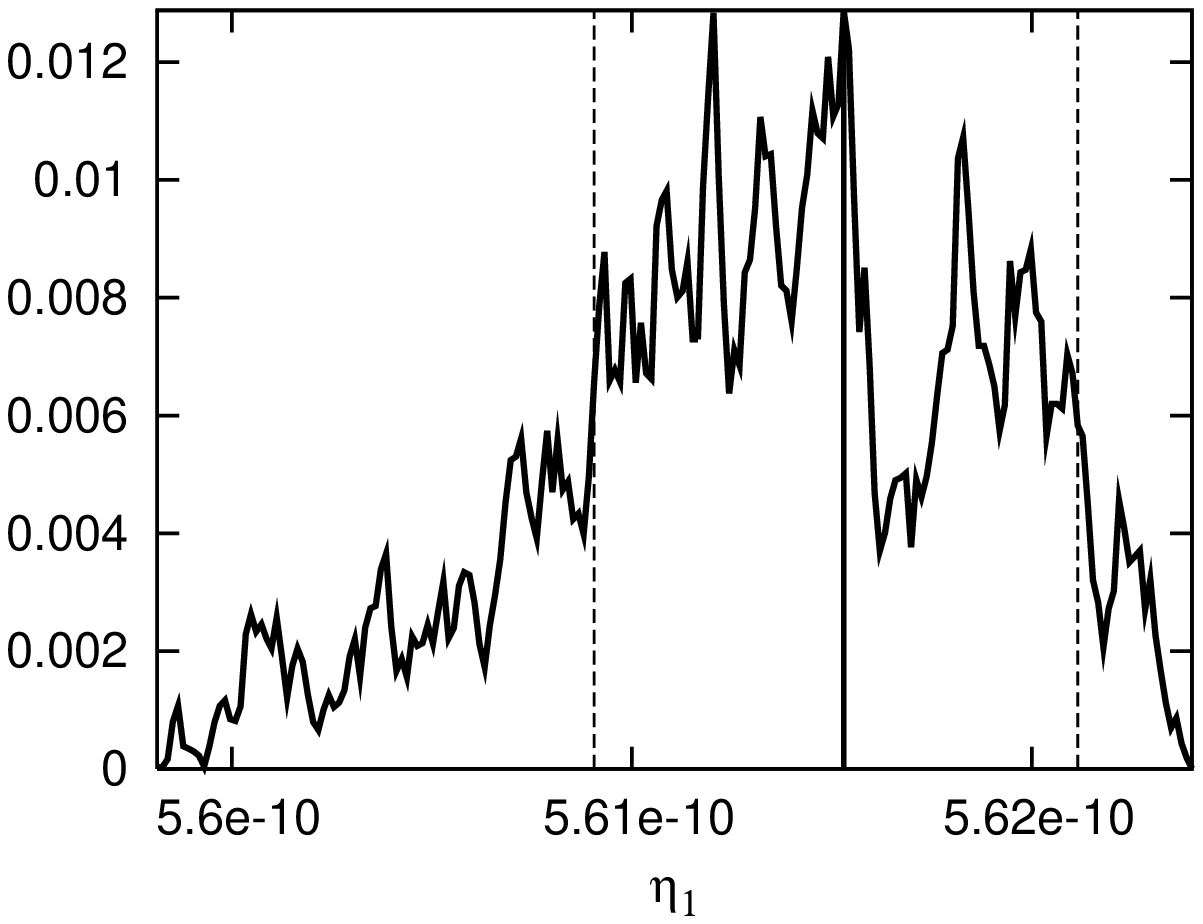}
\includegraphics[angle=270,scale=0.3]{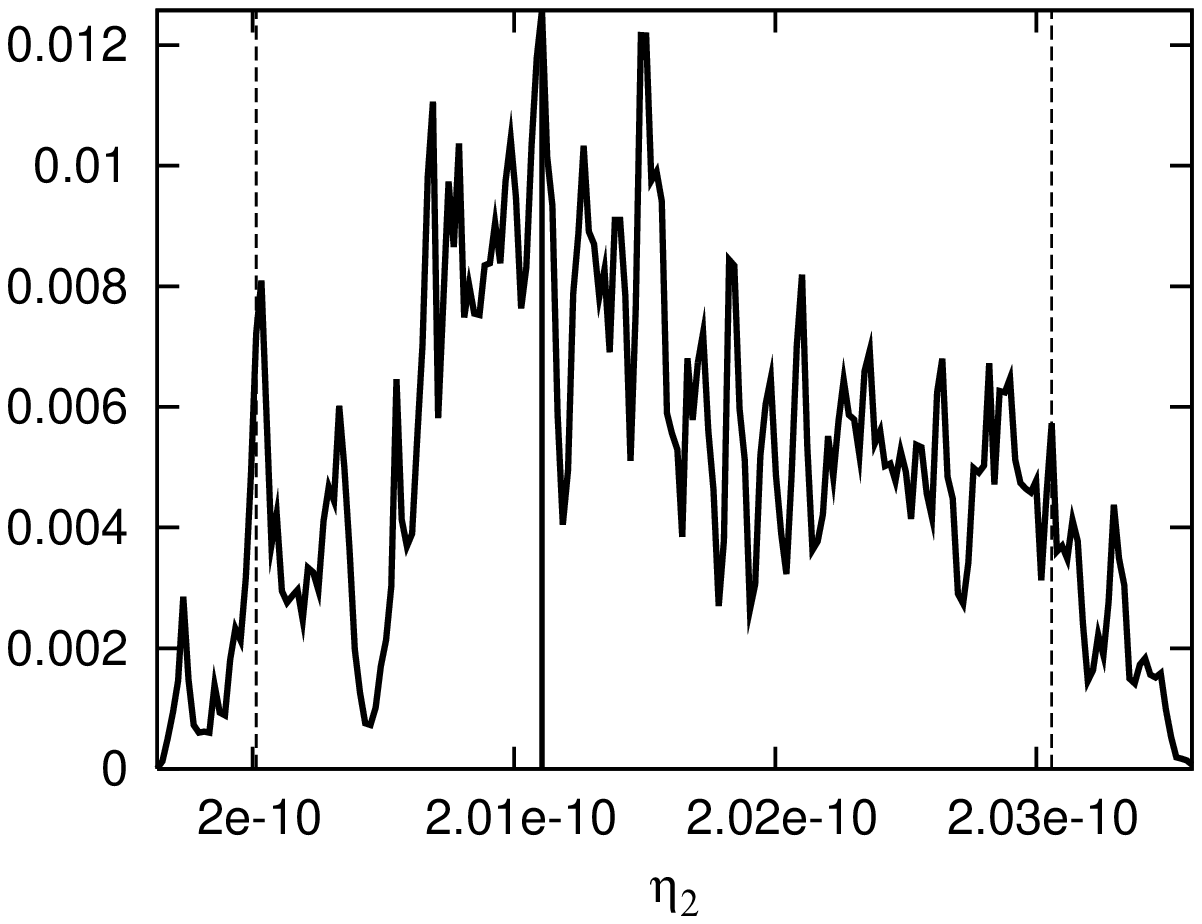}
\includegraphics[angle=270,scale=0.3]{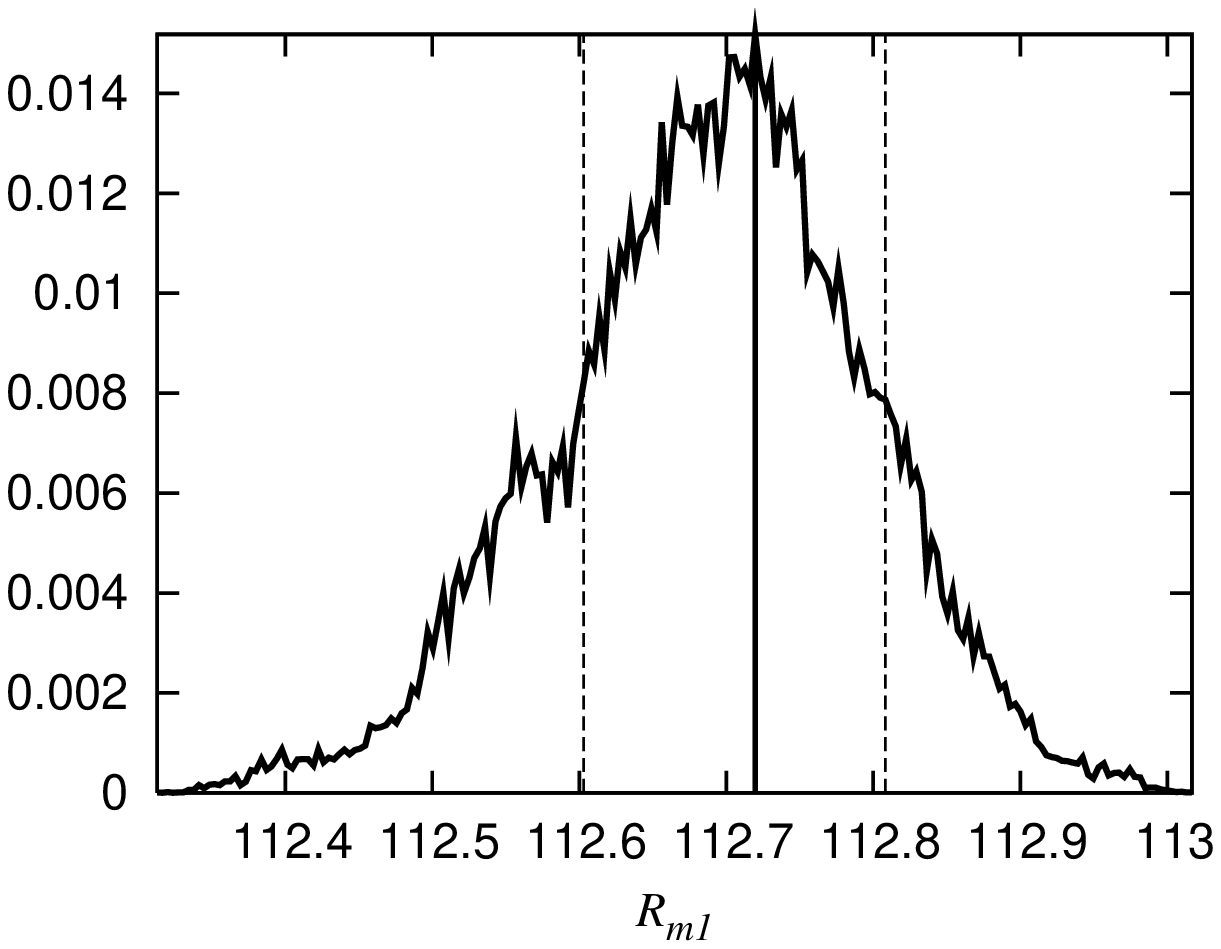}
\includegraphics[angle=270,scale=0.3]{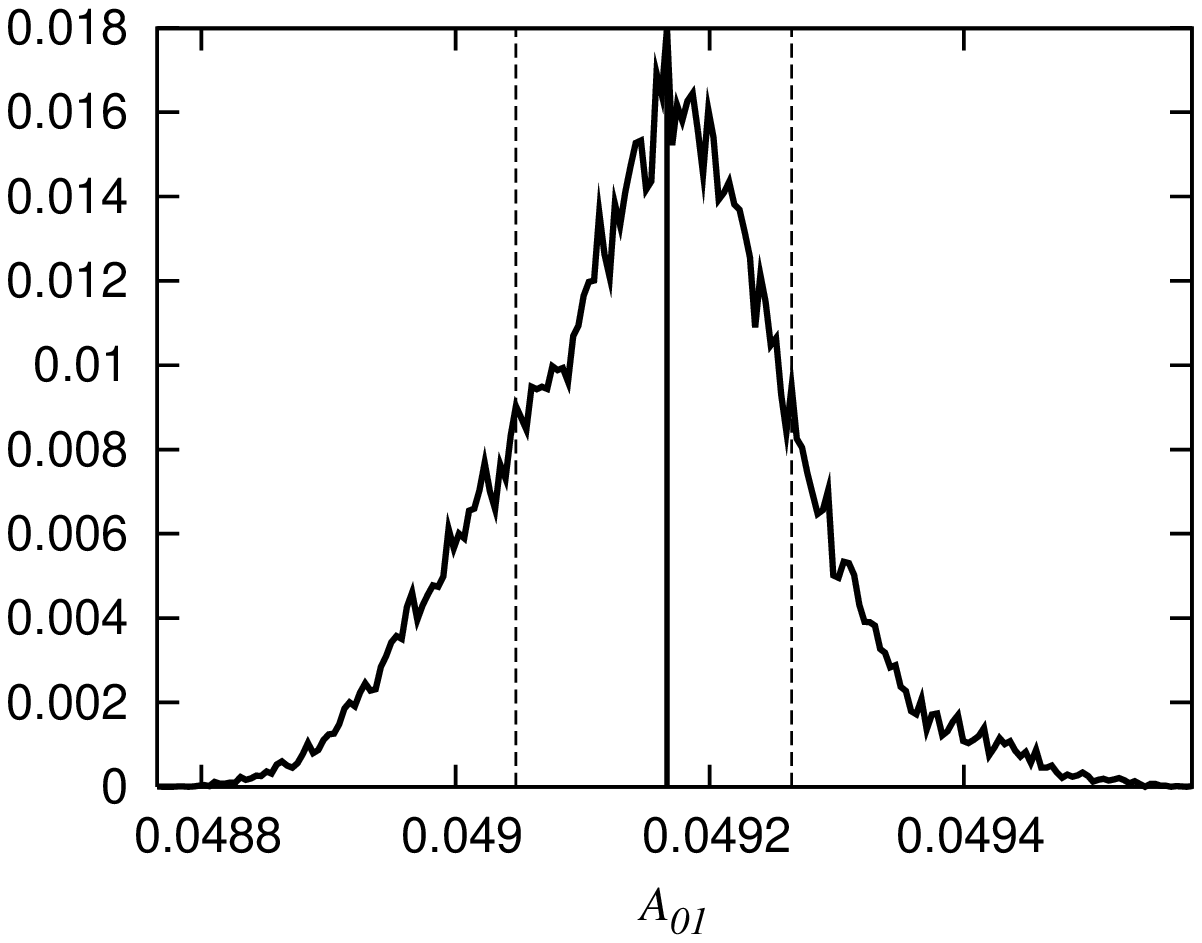}
\includegraphics[angle=270,scale=0.3]{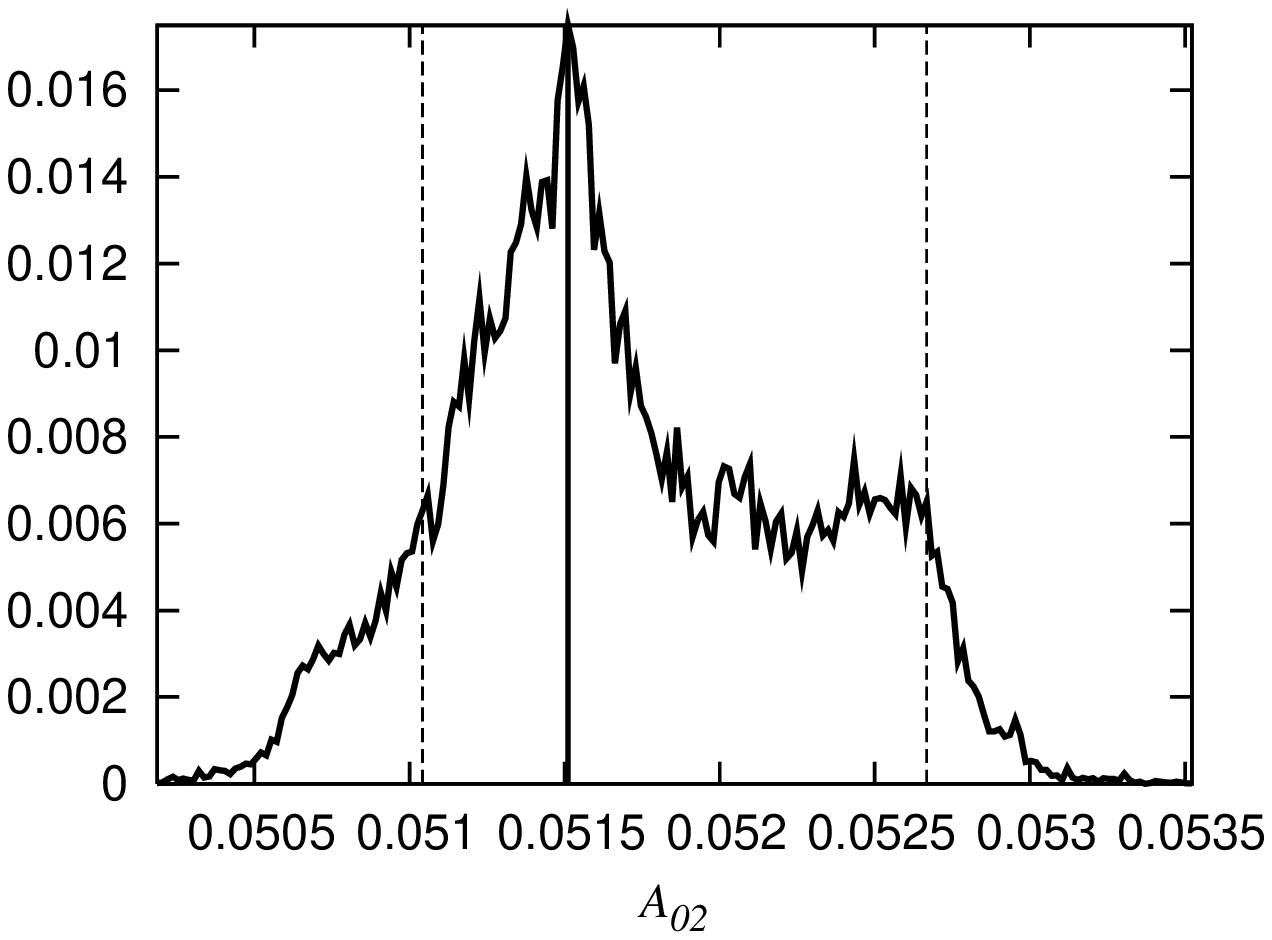}
\end{center}
\caption{The marginalized likelihood probability distributions of our 18-parameter 606
MCMC chains for each of the fitted parameters. The best-fit value (the peak of the distribution)
and the 68\% credible regions are given by the solid and dashed vertical lines.}\label{FigMCMC}
\end{figure*}

\begin{figure*}[tbp]
\begin{center}
\includegraphics[angle=270,scale=0.7]{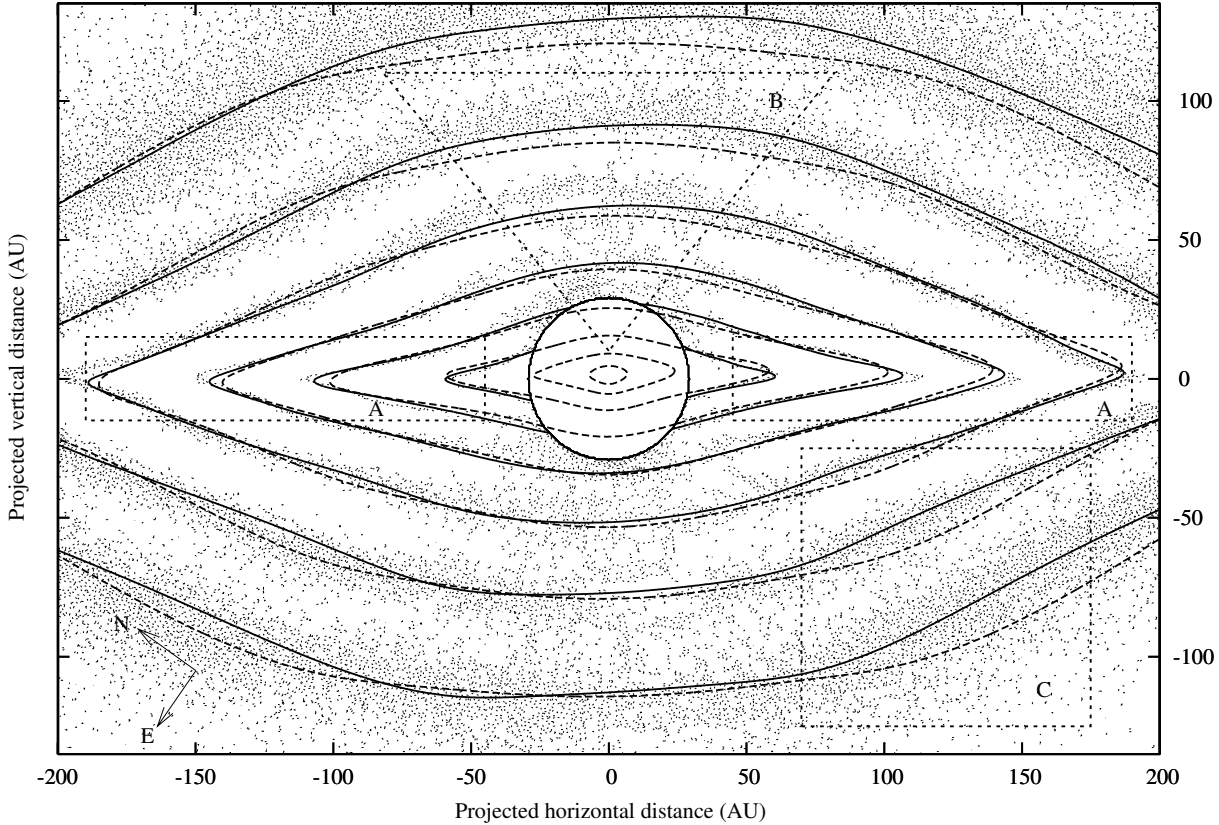}
\end{center}
\caption{Isophotal map of the light scattered by the dust orbiting $\beta$ Pic (dotted regions). Also shown are the single (dashed line) and the two-disk (solid line) model scattered light profiles. The brightness isophotes are quoted relative to $\beta Pic$, with the isophotes decreasing from $1\times10^{-6}$ $L_{disk}/L_{\beta Pic}$ with interval spacing equal to 0.5 in log base 10 units. From the overall low reduced $\chi^2=1.84$ of a single disk fit, one expects a reasonable fit for the observations. This is evident in the isophotal maps. The improvements offered by the two-disk model are shown in several regions: along the spine (A), forward scattering areas originating from larger scale heights (B) and the overall improvement in tracing the scattered light isophotes (C).}   
\label{TOT9AND18}

\end{figure*}

After we obtained the best single disk (9 parameter) fit, we ran our fitting procedure using two disks (18 parameters). 
The best-fit results for the one and two-disk models are listed in Table \ref{NineEighteenTwenty} along
with the associated errors on each parameter - the MCMC probability distributions of the 18 parameter 606 fit
are presented in Figure \ref{FigMCMC}. The minimum reduced $\chi^2$ for the single and two-disk fits are 1.83 and 1.18 respectively. In addition to the improvement in the reduced $\chi^2$, we present other qualitative improvements.
In Figure \ref{TOT9AND18} we display  the isophotal maps of our best-fit models, 
which shows that the two-disk model's isophotes better trace the observed isophotes.
The isophotal maps also reveal that two disks with axisymmetric dust distribution and non-isotropic scattering profiles can produce asymmetric brightness 
contours if they are inclined to each other and the line-of-sight (LOS) at non-zero angles. 
Degree of this asymmetry is further discussed in section 4.1.

\begin{figure*}[tbp]
\begin{center}
\includegraphics[angle=270,scale=0.7]{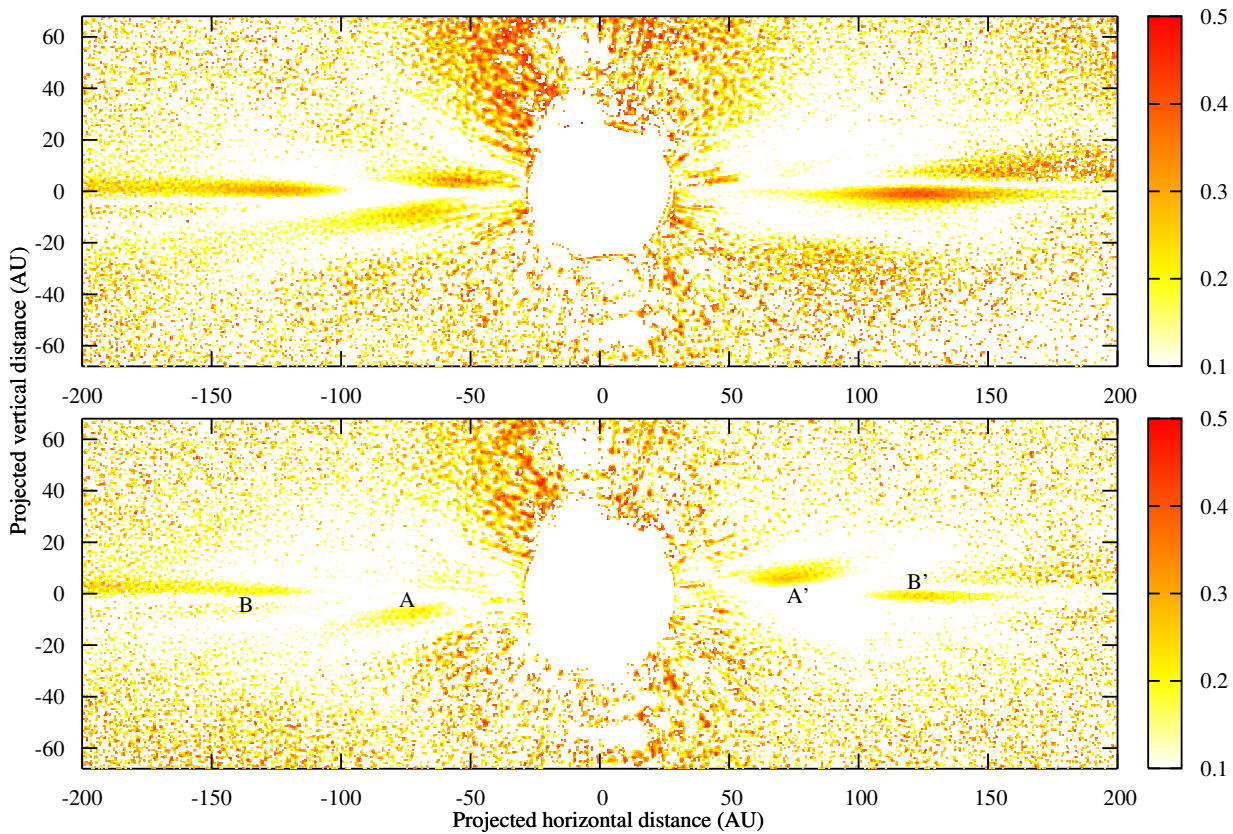}
\end{center}
\caption{The residual of the 606 ACS data following the removal of the scattered light produced by the single (top panel) and the 2 disk (bottom panel) models, presented relative to that of the composite disk. The second residual is not only cleaner, but is also symmetric following a $180\degr$ rotation. The identified ring-like structures (A and A', B and B') are similar to those identified by \citet{wah03}, albeit we are not certain whether their origin is real or failure of the model to properly describe the dust distribution.}
\label{TOTRES9AND18}
\end{figure*}

\begin{figure*}[tbp]
\begin{center}
\includegraphics[angle=270,scale=0.7]{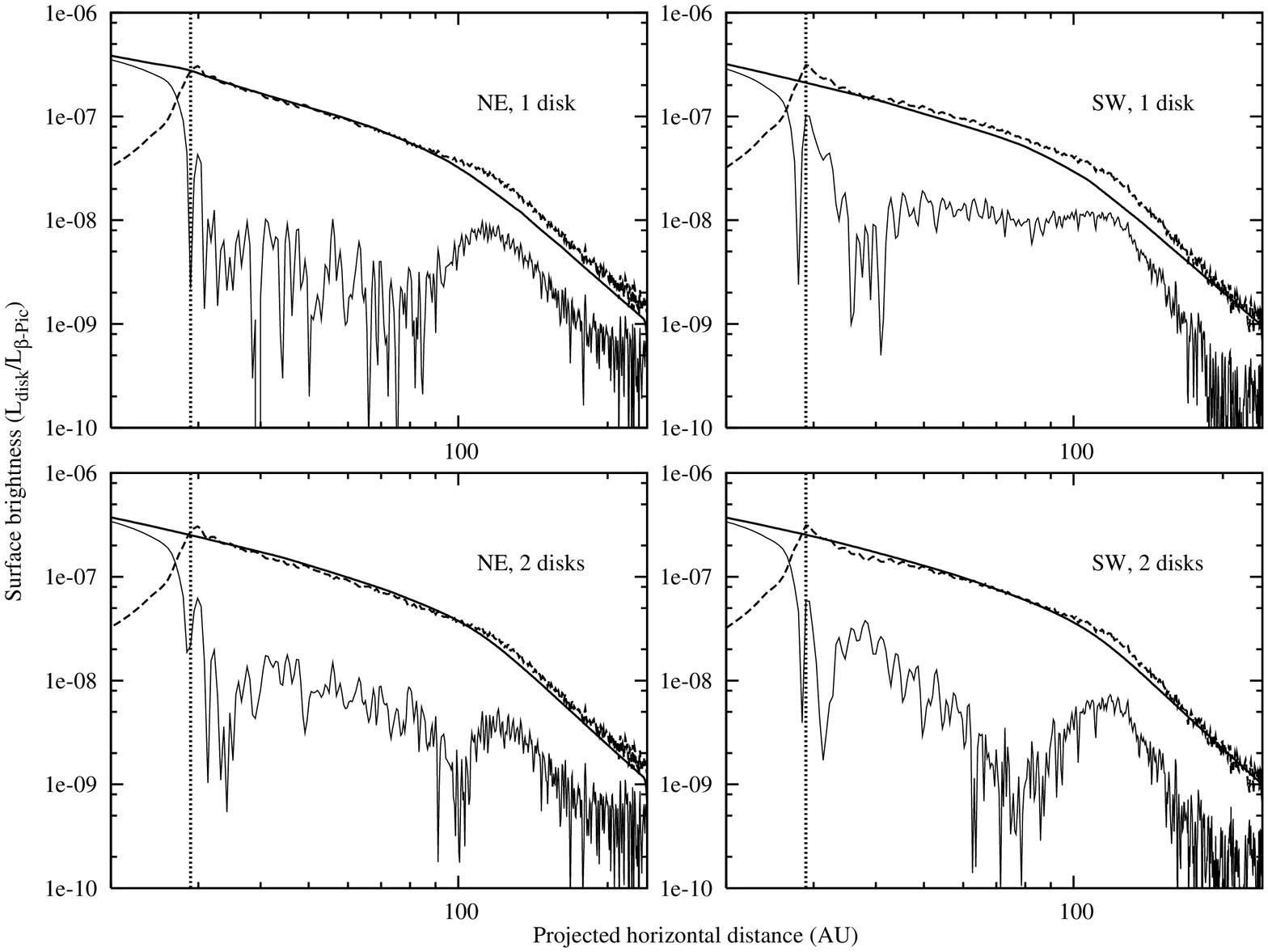}
\end{center}
\caption{Scattered light profile along the semimajor projection axis (the x-axis of Fig. \ref{TOT9AND18}). Top panels show the northeast and the southwest extensions (dashed lined) along with profiles produced by single disk model (thick solid line) and the residual (thin solid line) of these two. Bottom panels show the same for the two-disk model. Improvements in the fit can be seen by comparing the two-disk fit with the 1 disk fit. The dotted vertical line at 30\,AU represents the extent of the coronographic mask.}
\label{SPINE9AND18}
\end{figure*}

The total residuals following the removal of the one and then two-disk models from the data are presented in Figure \ref{TOTRES9AND18}.
Aside from producing overall smaller residuals, the two-disk model residuals are also symmetric, revealing
the main warp at (65-85)\,AU as well as an additional structure at (110-150)\,AU which is similar to
the rings observed by \citet{wah03}. We further discuss these structures in section 4.1.
In Figure \ref{SPINE9AND18} we present brightness
profile cuts along both the NE and the SW extensions of the composite disk, demonstrating once again the improvements gained from employing the two-disk model. 
For these reasons we believe the $\beta$ Pic disk is much more accurately described by our two axisymmetric disk model\footnote[1]{Bayesian model 
comparison 
over the one-disk model while taking into accounts the extra degrees of freedom of the two-disk model. It was was not applied here due to the clear quantitative and qualitative improvements of the two-disk model over the one-disk model}.
 
 \begin{deluxetable*}{ccccccc}
\tabletypesize{\scriptsize}
\tablecaption{9, 18 and 20 parameter fits}
\tablewidth{0pt}
\tablehead{
Parameter 		& 9 parameter     						& 9 parameter 606 														& 18 parameter    							& 18 parameter 606 														& 20 parameter    						& 20 parameter 606 \\
          		& 606 $\chi^{2}$ minimum 					& MCMC best-fit  														& 606 $\chi^{2}$ minimum 						& MCMC best-fit 														& 606 $\chi^{2}$ minimum 					& MCMC best-fit	\\
}
\startdata
$\chi^{2}$ 		& \chiNSOSmin $\times 10^{\chiNSOSEXPONENT}$ 			& \chiNSOSMCMCmin$^{+\chiNSOSMCMCplus}_{-\chiNSOSMCMCminus}$ $\times 10^{\chiNSOSEXPONENT}$ 					& \chiSOSmin $\times 10^{\chiSOSEXPONENT}$ 				& \chiSOSMCMCmin$^{+\chiSOSMCMCplus}_{-\chiSOSMCMCminus}$ $\times 10^{\chiSOSEXPONENT}$ 					& \chiTSOSmin $\times 10^{\chiTSOSEXPONENT}$			& \chiTSOSMCMCmin$^{+\chiTSOSMCMCplus}_{-\chiTSOSMCMCminus}$ $\times 10^{\chiTSOSEXPONENT}$			\\ 
Reduced $\chi^{2}$      & \chiReducedNSOSmin & \chiReducedNSOSMCMCmin$^{+\chiReducedNSOSMCMCplus}_{-\chiReducedNSOSMCMCminus}$ & \chiReducedSOSmin                                                     & \chiReducedSOSMCMCmin$^{+\chiReducedSOSMCMCplus}_{-\chiReducedSOSMCMCminus}$ & \chiReducedTSOSmin                                            & \chiReducedTSOSMCMCmin$^{+\chiReducedTSOSMCMCplus}_{-\chiReducedTSOSMCMCminus}$ \\
$i_{1}$ ($^{o}$)	& \paramONENSOSmin 						& \paramONENSOSMCMCmin$^{+\paramONENSOSMCMCplus}_{-\paramONENSOSMCMCminus}$ 							& \paramONESOSmin 							& \paramONESOSMCMCmin$^{+\paramONESOSMCMCplus}_{-\paramONESOSMCMCminus}$ 							& \paramONETSOSmin						& \paramONETSOSMCMCmin$^{+\paramONETSOSMCMCplus}_{-\paramONETSOSMCMCminus}$\\
$p_{1}$ 		& \paramTWONSOSmin 						& \paramTWONSOSMCMCmin$^{+\paramTWONSOSMCMCplus}_{-\paramTWONSOSMCMCminus}$ 							& \paramTWOSOSmin 							& \paramTWOSOSMCMCmin$^{+\paramTWOSOSMCMCplus}_{-\paramTWOSOSMCMCminus}$ 							& \paramTWOTSOSmin						& \paramTWOTSOSMCMCmin$^{+\paramTWOTSOSMCMCplus}_{-\paramTWOTSOSMCMCminus}$ \\
$\phi_{1}$ ($^{o}$) 	& \paramTHREENSOSmin 						& \paramTHREENSOSMCMCmin$^{+\paramTHREENSOSMCMCplus}_{-\paramTHREENSOSMCMCminus}$ 						& \paramTHREESOSmin  							& \paramTHREESOSMCMCmin$^{+\paramTHREESOSMCMCplus}_{-\paramTHREESOSMCMCminus}$ 					 		& \paramTHREETSOSmin 						& \paramTHREETSOSMCMCmin$^{+\paramTHREETSOSMCMCplus}_{-\paramTHREETSOSMCMCminus}$ \\
$\alpha_{1}$ 		& \paramFOURNSOSmin 						& \paramFOURNSOSMCMCmin$^{+\paramFOURNSOSMCMCplus}_{-\paramFOURNSOSMCMCminus}$ 							& \paramFOURSOSmin 							& \paramFOURSOSMCMCmin$^{+\paramFOURSOSMCMCplus}_{-\paramFOURSOSMCMCminus}$ 							& \paramFOURTSOSmin						& \paramFOURTSOSMCMCmin$^{+\paramFOURTSOSMCMCplus}_{-\paramFOURTSOSMCMCminus}$ 		\\
$\beta_{1}$ 		& \paramFIVENSOSmin 						& \paramFIVENSOSMCMCmin$^{+\paramFIVENSOSMCMCplus}_{-\paramFIVENSOSMCMCminus}$ 							& \paramFIVESOSmin 							& \paramFIVESOSMCMCmin$^{+\paramFIVESOSMCMCplus}_{-\paramFIVESOSMCMCminus}$ 							& \paramFIVETSOSmin						& \paramFIVETSOSMCMCmin$^{+\paramFIVETSOSMCMCplus}_{-\paramFIVETSOSMCMCminus}$ 		\\
$\gamma_{1}$ 		& \paramSIXNSOSmin 						& \paramSIXNSOSMCMCmin$^{+\paramSIXNSOSMCMCplus}_{-\paramSIXNSOSMCMCminus}$ 							& \paramSIXSOSmin							& \paramSIXSOSMCMCmin$^{+\paramSIXSOSMCMCplus}_{-\paramSIXSOSMCMCminus}$ 							& \paramSIXTSOSmin						& \paramSIXTSOSMCMCmin$^{+\paramSIXTSOSMCMCplus}_{-\paramSIXTSOSMCMCminus}$ 		\\
$i_{2}$	($^{o}$) 	& n/a								& n/a																& \paramSEVENSOSmin							& \paramSEVENSOSMCMCmin$^{+\paramSEVENSOSMCMCplus}_{-\paramSEVENSOSMCMCminus}$ 							& \paramSEVENTSOSmin						& \paramSEVENTSOSMCMCmin$^{+\paramSEVENTSOSMCMCplus}_{-\paramSEVENTSOSMCMCminus}$ 		\\
$p_{2}$			& n/a								& n/a																& \paramEIGHTSOSmin							& \paramEIGHTSOSMCMCmin$^{+\paramEIGHTSOSMCMCplus}_{-\paramEIGHTSOSMCMCminus}$ 							& \paramEIGHTTSOSmin						& \paramEIGHTTSOSMCMCmin$^{+\paramEIGHTTSOSMCMCplus}_{-\paramEIGHTTSOSMCMCminus}$ 		\\
$\phi_{2}$ ($^{o}$) 	& n/a 								& n/a 																& \paramNINESOSmin							& \paramNINESOSMCMCmin$^{+\paramNINESOSMCMCplus}_{-\paramNINESOSMCMCminus}$							& \paramNINETSOSmin						& \paramNINETSOSMCMCmin$^{+\paramNINETSOSMCMCplus}_{-\paramNINETSOSMCMCminus}$ 		\\
$\alpha_{2}$ 		& n/a 								& n/a 																& \paramTENSOSmin 							& \paramTENSOSMCMCmin$^{+\paramTENSOSMCMCplus}_{-\paramTENSOSMCMCminus}$ 							& \paramTENTSOSmin						& \paramTENTSOSMCMCmin$^{+\paramTENTSOSMCMCplus}_{-\paramTENTSOSMCMCminus}$ 		\\
$\beta_{2}$ 		& n/a 								& n/a 																& \paramTENONESOSmin 							& \paramTENONESOSMCMCmin$^{+\paramTENONESOSMCMCplus}_{-\paramTENONESOSMCMCminus}$ 						& \paramTENONETSOSmin						& \paramTENONETSOSMCMCmin$^{+\paramTENONETSOSMCMCplus}_{-\paramTENONETSOSMCMCminus}$ 	\\
$\gamma_{2}$ 		& n/a 								& n/a 																& \paramTENTWOSOSmin 							& \paramTENTWOSOSMCMCmin$^{+\paramTENTWOSOSMCMCplus}_{-\paramTENTWOSOSMCMCminus}$ 						& \paramTENTWOTSOSmin						& \paramTENTWOTSOSMCMCmin$^{+\paramTENTWOTSOSMCMCplus}_{-\paramTENTWOTSOSMCMCminus}$ 	\\
$R_{m1}$ (AU) 		& \paramSEVENNSOSmin		 				& \paramSEVENNSOSMCMCmin$^{+\paramSEVENNSOSMCMCplus}_{-\paramSEVENNSOSMCMCminus}$ 						& \paramTENSIXSOSmin 							& \paramTENSIXSOSMCMCmin$^{+\paramTENSIXSOSMCMCplus}_{-\paramTENSIXSOSMCMCminus}$ 						& \paramTENSIXTSOSmin						& \paramTENSIXTSOSMCMCmin$^{+\paramTENSIXTSOSMCMCplus}_{-\paramTENSIXTSOSMCMCminus}$ 	\\
$R_{m2}$ (AU) 		& n/a 								& n/a 																& \paramTENTHREESOSmin 							& \paramTENTHREESOSMCMCmin$^{+\paramTENTHREESOSMCMCplus}_{-\paramTENTHREESOSMCMCminus}$ 					& \paramTENTHREETSOSmin						& \paramTENTHREETSOSMCMCmin$^{+\paramTENTHREETSOSMCMCplus}_{-\paramTENTHREETSOSMCMCminus}$ \\
$\eta_{1}$ 		& \paramEIGHTNSOSmin $\times 10^{\paramEIGHTNSOSEXPONENT}$	& \paramEIGHTNSOSMCMCmin$^{+\paramEIGHTNSOSMCMCplus}_{-\paramEIGHTNSOSMCMCminus}$ $\times 10^{\paramEIGHTNSOSEXPONENT}$		& \paramTENFOURSOSmin $\times 10^{\paramTENFOURSOSEXPONENT}$ 		& \paramTENFOURSOSMCMCmin$^{+\paramTENFOURSOSMCMCplus}_{-\paramTENFOURSOSMCMCminus}$ $\times 10^{\paramTENFOURSOSEXPONENT}$ 	& \paramTENFOURTSOSmin	$\times 10^{\paramTENFOURTSOSEXPONENT}$	& \paramTENFOURTSOSMCMCmin$^{+\paramTENFOURTSOSMCMCplus}_{-\paramTENFOURTSOSMCMCminus}$ $\times 10^{\paramTENFOURTSOSEXPONENT}$ 	\\
$\eta_{2}$ 		& n/a 								& n/a 																& \paramTENFIVESOSmin $\times 10^{\paramTENFIVESOSEXPONENT}$		& \paramTENFIVESOSMCMCmin$^{+\paramTENFIVESOSMCMCplus}_{-\paramTENFIVESOSMCMCminus}$ $\times 10^{\paramTENFIVESOSEXPONENT}$	& \paramTENFIVETSOSmin	$\times 10^{\paramTENFIVETSOSEXPONENT}$	& \paramTENFIVETSOSMCMCmin$^{+\paramTENFIVETSOSMCMCplus}_{-\paramTENFIVETSOSMCMCminus}$ $\times 10^{\paramTENFIVETSOSEXPONENT}$	\\
$A_{01}$ 		& \paramNINENSOSmin 						& \paramNINENSOSMCMCmin$^{+\paramNINENSOSMCMCplus}_{-\paramNINENSOSMCMCminus}$ 							& \paramTENSEVENSOSmin 							& \paramTENSEVENSOSMCMCmin$^{+\paramTENSEVENSOSMCMCplus}_{-\paramTENSEVENSOSMCMCminus}$ 					& \paramTENSEVENTSOSmin						& \paramTENSEVENTSOSMCMCmin$^{+\paramTENSEVENTSOSMCMCplus}_{-\paramTENSEVENTSOSMCMCminus}$\\
$A_{02}$ 		& n/a 								& n/a 																& \paramTENEIGHTSOSmin 							& \paramTENEIGHTSOSMCMCmin$^{+\paramTENEIGHTSOSMCMCplus}_{-\paramTENEIGHTSOSMCMCminus}$ 					& \paramTENEIGHTTSOSmin						& \paramTENEIGHTTSOSMCMCmin$^{+\paramTENEIGHTTSOSMCMCplus}_{-\paramTENEIGHTTSOSMCMCminus}$\\
$g1$			& n/a								& n/a																& n/a									& n/a																& \paramTENNINETSOSmin						& \paramTENNINETSOSMCMCmin$^{+\paramTENNINETSOSMCMCplus}_{-\paramTENNINETSOSMCMCminus}$\\
$g2$			& n/a								& n/a																& n/a									& n/a																& \paramTENTENTSOSmin						& \paramTENTENTSOSMCMCmin$^{+\paramTENTENTSOSMCMCplus}_{-\paramTENTENTSOSMCMCminus}$\\
\label{NineEighteenTwenty}
\enddata
\tablenotetext{a}{As discussed in $\S$\ref{SecLargeNumberOfDataPoints} the formal errors have been scaled up by a factor of 10.}
\end{deluxetable*}

\begin{figure}[tbp]
\begin{center}
\includegraphics[angle=270,scale=0.7]{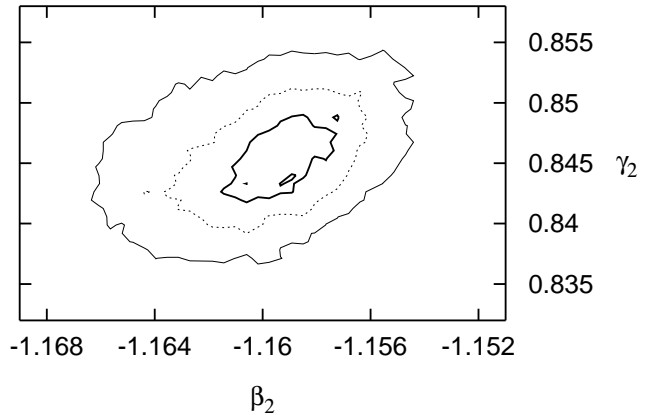}
\end{center}
\caption{A plot of the correlation between the radial index ($\beta_{2}$) and the flare index ($\gamma_{2}$) displaying the 
33\% (solid thick curve), 68\% (dotted curve) and the 95\% (solid thin curve) credible regions for the 18-parameter 606 ACS data. While some correlation between these parameters is expected as they both describe light scattered at larger radii, the two are still well constrained.}
\label{COR}
\end{figure}

We also investigate for possible correlations between the various parameters of our model.
The most significant correlation is between the secondary's outer radial index ($\beta_{2}$)
and the flare index ($\gamma_{2}$).
This correlation is shown in Figure \ref{COR} for the 18-parameter F606W data set. Other parameters were not found to be notably correlated.

\subsection{Chromatic dependence of modeled parameters}

Prior to the \citet{gol06} observations of $\beta$ Pic, it was widely accepted that the orbiting dust was composed of neutrally scattering grains 
with sizes larger than several microns (ex. \citealt{les93}). With the higher photometric precision of the ACS, \citet{gol06} analyzed the disk's 
colors with three filters [a narrow B-filter (435 nm), a broadbard V (606 nm), and a broadband I (814 nm)] and concluded that this assumption was wrong. 
Complicating matters further, their observations showed asymmetries in the flux ratios taken at different wavebands about the projected major and 
minor axis of the disk, which imposes additional constrains on the models and suggests that a true description of the dust distribution will be 
incomplete until wavelength-dependent models are developed.

Unfortunately, our multi-parametric dust distribution model is not wavelength-dependent and neither are the phase functions we use\footnote{zodiacal light function is derived from particles scattering sunlight which are much larger than the wavelengths they scatter and thus one can think of it as color-neutral scattering function}. In addition, any modifications would introduce additional parameters and thus slow down the search through the parameter space. Considering these points and the already low reduced $\chi^2$ we have been able to obtain from our wavelength-independent model, we chose not to increase the complexity of the model. 

There are some benefits in our fitting the same model to data observed at different wavelengths.
Fitting to images of the disk at different wavelengths and comparing the derived parameters 
can reveal whether the scattering
in the $\beta$ Pic disk is wavelength dependent and even put some constrains on the 
 optical properties of dust. Conversely, assuming the scattering in the $\beta$ Pic disk to be 
color-neutral and comparing the results from fits using different wavelengths, we can obtain an independent estimate of the robustness of our fit.  

Table \ref{CHROMATIC} lists the best fits obtained for the three wavebands. Overall there is a good agreement between the parameters, especially for the values describing the primary. Several of the parameters describing the secondary [the inner radial index $\alpha_2$, radius of the power law break $R_{m2}$ and the scale height index $A_0$] show a much
wider range than those corresponding to the primary disk. For instance the apparent increase of $\alpha_2$ in the shorter wavelength
bands might suggest that there is substantial reddening in the inner part of the secondary. However, we also note that any errors in
the subtraction near the coronograhic mask would effect $\alpha_2$ the most and thus we believe that this result should be viewed with some skepticism.

\begin{figure}[tbp]
\begin{center}
\includegraphics[angle=270,scale=0.7]{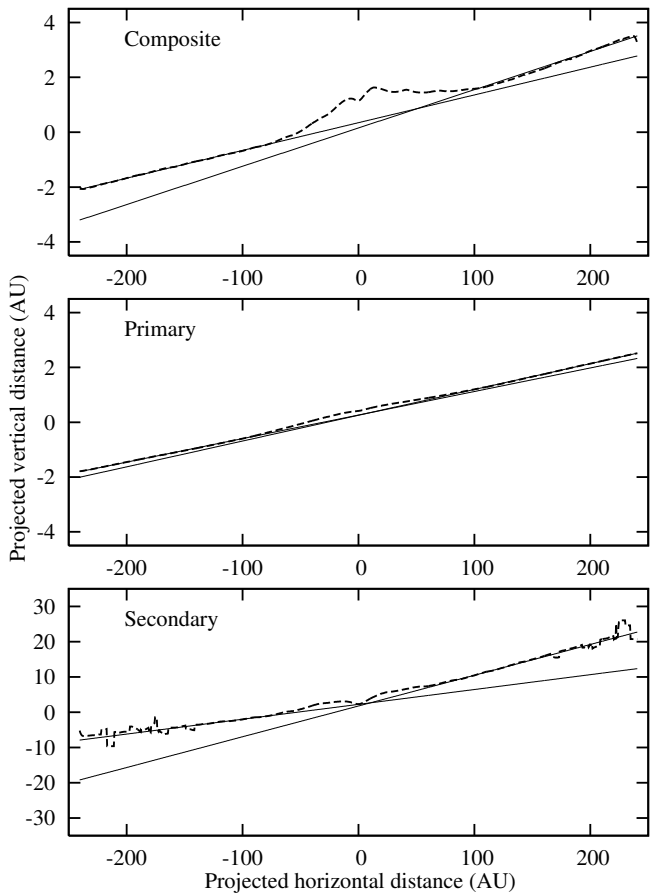}
\end{center}
\caption{Spines of the composite, primary and secondary disks (dashed lines) determined by fitting gaussians to the vertical profiles
	at each step along the semimajor axis. The solid lines represent linear fits to each extension of the disk,
	which we use to determine the inclination values of each component presented in Table \ref{SPINECPSTAB}. The solid lines are enlongated into
	the opposite extension for visibility.}
\label{SPINECPS}
\end{figure}

We also use the models produced for the three filters to constrain the inclination of the secondary to the primary. The fits for the locations of the spines of both disks (as well as the composite) are shown if Figure \ref{SPINECPS}. Table \ref{SPINECPSTAB} lists the corresponding inclinations of the NE and SW extensions measured counterclockwise to the major projection axis. The value for the angular separation is of particular interest since this value has been previously reported. Our result, $3.2\pm1.3\degr$, is consistent with the $\sim$5$\degr$ value reported by \citet{gol06}.

\begin{deluxetable*}{ccccccc}
\tabletypesize{\scriptsize}
\tablecaption{Chromatic variations}
\tablewidth{0pt}
\tablehead{
Parameter		& 606 $\chi^{2}$	& 606 MCMC	& 435 $\chi^{2}$	& 435 MCMC	& 814 $\chi^{2}$	& 814 MCMC \\
			& minimum               & best-fit	& minimum               & best-fit	& minimum               & best-fit\\}
\startdata
$\chi^{2}$    		& \chiSOSmin $\times 10^{\chiSOSEXPONENT}$			& \chiSOSMCMCmin$^{+\chiSOSMCMCplus}_{-\chiSOSMCMCminus}$ $\times 10^{\chiSOSEXPONENT}$						& \chiFTFmin	$\times 10^{\chiFTFEXPONENT}$			& \chiFTFMCMCmin$^{+\chiFTFMCMCplus}_{-\chiFTFMCMCminus}$ $\times 10^{\chiFTFEXPONENT}$						& \chiEOFmin	$\times 10^{\chiEOFEXPONENT}$			& \chiEOFMCMCmin$^{+\chiEOFMCMCplus}_{-\chiEOFMCMCminus}$ $\times 10^{\chiEOFEXPONENT}$	\\ 
Reduced $\chi^{2}$	& \chiReducedSOSmin 						& \chiReducedSOSMCMCmin$^{+\chiReducedSOSMCMCplus}_{-\chiReducedSOSMCMCminus}$ 							& \chiReducedFTFmin						& \chiReducedFTFMCMCmin$^{+\chiReducedFTFMCMCplus}_{-\chiReducedFTFMCMCminus}$ 							& \chiReducedEOFmin						& \chiReducedEOFMCMCmin$^{+\chiReducedEOFMCMCplus}_{-\chiReducedEOFMCMCminus}$ \\ 
$i_{1}$	($^{o}$)	& \paramONESOSmin 						& \paramONESOSMCMCmin$^{+\paramONESOSMCMCplus}_{-\paramONESOSMCMCminus}$ 							& \paramONEFTFmin 						& \paramONEFTFMCMCmin$^{+\paramONEFTFMCMCplus}_{-\paramONEFTFMCMCminus}$ 						 	& \paramONEEOFmin 						& \paramONEEOFMCMCmin$^{+\paramONEEOFMCMCplus}_{-\paramONEEOFMCMCminus}$ \\
$p_{1}$			& \paramTWOSOSmin 						& \paramTWOSOSMCMCmin$^{+\paramTWOSOSMCMCplus}_{-\paramTWOSOSMCMCminus}$ 							& \paramTWOFTFmin						& \paramTWOFTFMCMCmin$^{+\paramTWOFTFMCMCplus}_{-\paramTWOFTFMCMCminus}$ 							& \paramTWOEOFmin						& \paramTWOEOFMCMCmin$^{+\paramTWOEOFMCMCplus}_{-\paramTWOEOFMCMCminus}$\\
$\phi_{1}$ ($^{o}$)	& \paramTHREESOSmin 						& \paramTHREESOSMCMCmin$^{+\paramTHREESOSMCMCplus}_{-\paramTHREESOSMCMCminus}$ 							& \paramTHREEFTFmin 						& \paramTHREEFTFMCMCmin$^{+\paramTHREEFTFMCMCplus}_{-\paramTHREEFTFMCMCminus}$ 							& \paramTHREEEOFmin						& \paramTHREEEOFMCMCmin$^{+\paramTHREEEOFMCMCplus}_{-\paramTHREEEOFMCMCminus}$\\
$\alpha_{1}$		& \paramFOURSOSmin 						& \paramFOURSOSMCMCmin$^{+\paramFOURSOSMCMCplus}_{-\paramFOURSOSMCMCminus}$ 							& \paramFOURFTFmin						& \paramFOURFTFMCMCmin$^{+\paramFOURFTFMCMCplus}_{-\paramFOURFTFMCMCminus}$ 							& \paramFOUREOFmin						& \paramFOUREOFMCMCmin$^{+\paramFOUREOFMCMCplus}_{-\paramFOUREOFMCMCminus}$\\
$\beta_{1}$		& \paramFIVESOSmin						& \paramFIVESOSMCMCmin$^{+\paramFIVESOSMCMCplus}_{-\paramFIVESOSMCMCminus}$ 							& \paramFIVEFTFmin						& \paramFIVEFTFMCMCmin$^{+\paramFIVEFTFMCMCplus}_{-\paramFIVEFTFMCMCminus}$ 							& \paramFIVEEOFmin						& \paramFIVEEOFMCMCmin$^{+\paramFIVEEOFMCMCplus}_{-\paramFIVEEOFMCMCminus}$\\
$\gamma_{1}$		& \paramSIXSOSmin 						& \paramSIXSOSMCMCmin$^{+\paramSIXSOSMCMCplus}_{-\paramSIXSOSMCMCminus}$ 							& \paramSIXFTFmin						& \paramSIXFTFMCMCmin$^{+\paramSIXFTFMCMCplus}_{-\paramSIXFTFMCMCminus}$ 							& \paramSIXEOFmin						& \paramSIXEOFMCMCmin$^{+\paramSIXEOFMCMCplus}_{-\paramSIXEOFMCMCminus}$\\
$i_{2}$	($^{o}$)	& \paramSEVENSOSmin 						& \paramSEVENSOSMCMCmin$^{+\paramSEVENSOSMCMCplus}_{-\paramSEVENSOSMCMCminus}$ 							& \paramSEVENFTFmin						& \paramSEVENFTFMCMCmin$^{+\paramSEVENFTFMCMCplus}_{-\paramSEVENFTFMCMCminus}$ 							& \paramSEVENEOFmin						& \paramSEVENEOFMCMCmin$^{+\paramSEVENEOFMCMCplus}_{-\paramSEVENEOFMCMCminus}$\\
$p_{2}$			& \paramEIGHTSOSmin 						& \paramEIGHTSOSMCMCmin$^{+\paramEIGHTSOSMCMCplus}_{-\paramEIGHTSOSMCMCminus}$ 							& \paramEIGHTFTFmin						& \paramEIGHTFTFMCMCmin$^{+\paramEIGHTFTFMCMCplus}_{-\paramEIGHTFTFMCMCminus}$			 				& \paramEIGHTEOFmin						& \paramEIGHTEOFMCMCmin$^{+\paramEIGHTEOFMCMCplus}_{-\paramEIGHTEOFMCMCminus}$\\
$\phi_{2}$ ($^{o}$)	& \paramNINESOSmin 						& \paramNINESOSMCMCmin$^{+\paramNINESOSMCMCplus}_{-\paramNINESOSMCMCminus}$ 							& \paramNINEFTFmin						& \paramNINEFTFMCMCmin$^{+\paramNINEFTFMCMCplus}_{-\paramNINEFTFMCMCminus}$ 							& \paramNINEEOFmin						& \paramNINEEOFMCMCmin$^{+\paramNINEEOFMCMCplus}_{-\paramNINEEOFMCMCminus}$\\
$\alpha_{2}$		& \paramTENSOSmin 						& \paramTENSOSMCMCmin$^{+\paramTENSOSMCMCplus}_{-\paramTENSOSMCMCminus}$ 							& \paramTENFTFmin						& \paramTENFTFMCMCmin$^{+\paramTENFTFMCMCplus}_{-\paramTENFTFMCMCminus}$ 							& \paramTENEOFmin						& \paramTENEOFMCMCmin$^{+\paramTENEOFMCMCplus}_{-\paramTENEOFMCMCminus}$\\
$\beta_{2}$		& \paramTENONESOSmin 						& \paramTENONESOSMCMCmin$^{+\paramTENONESOSMCMCplus}_{-\paramTENONESOSMCMCminus}$ 						& \paramTENONEFTFmin						& \paramTENONEFTFMCMCmin$^{+\paramTENONEFTFMCMCplus}_{-\paramTENONEFTFMCMCminus}$					 	& \paramTENONEEOFmin						& \paramTENONEEOFMCMCmin$^{+\paramTENONEEOFMCMCplus}_{-\paramTENONEEOFMCMCminus}$\\
$\gamma_{2}$		& \paramTENTWOSOSmin 						& \paramTENTWOSOSMCMCmin$^{+\paramTENTWOSOSMCMCplus}_{-\paramTENTWOSOSMCMCminus}$ 						& \paramTENTWOFTFmin						& \paramTENTWOFTFMCMCmin$^{+\paramTENTWOFTFMCMCplus}_{-\paramTENTWOFTFMCMCminus}$ 						& \paramTENTWOEOFmin						& \paramTENTWOEOFMCMCmin$^{+\paramTENTWOEOFMCMCplus}_{-\paramTENTWOEOFMCMCminus}$\\
$R_{m1}$ (AU)		& \paramTENSIXSOSmin 						& \paramTENSIXSOSMCMCmin$^{+\paramTENSIXSOSMCMCplus}_{-\paramTENSIXSOSMCMCminus}$ 						& \paramTENSIXFTFmin						& \paramTENSIXFTFMCMCmin$^{+\paramTENSIXFTFMCMCplus}_{-\paramTENSIXFTFMCMCminus}$ 						& \paramTENSIXEOFmin						& \paramTENSIXEOFMCMCmin$^{+\paramTENSIXEOFMCMCplus}_{-\paramTENSIXEOFMCMCminus}$\\
$R_{m2}$ (AU)		& \paramTENTHREESOSmin 						& \paramTENTHREESOSMCMCmin$^{+\paramTENTHREESOSMCMCplus}_{-\paramTENTHREESOSMCMCminus}$ 			 		& \paramTENTHREEFTFmin						& \paramTENTHREEFTFMCMCmin$^{+\paramTENTHREEFTFMCMCplus}_{-\paramTENTHREEFTFMCMCminus}$						& \paramTENTHREEEOFmin						& \paramTENTHREEEOFMCMCmin$^{+\paramTENTHREEEOFMCMCplus}_{-\paramTENTHREEEOFMCMCminus}$\\
$\eta_{1}$		& \paramTENFOURSOSmin $\times 10^{\paramTENFOURSOSEXPONENT}$	& \paramTENFOURSOSMCMCmin$^{+\paramTENFOURSOSMCMCplus}_{-\paramTENFOURSOSMCMCminus}$ $\times 10^{\paramTENFOURSOSEXPONENT}$ 	& \paramTENFOURFTFmin $\times 10^{\paramTENFOURFTFEXPONENT}$	& \paramTENFOURFTFMCMCmin$^{+\paramTENFOURFTFMCMCplus}_{-\paramTENFOURFTFMCMCminus}$ $\times 10^{\paramTENFOURFTFEXPONENT}$	& \paramTENFOUREOFmin $\times 10^{\paramTENFOUREOFEXPONENT}$	& \paramTENFOUREOFMCMCmin$^{+\paramTENFOUREOFMCMCplus}_{-\paramTENFOUREOFMCMCminus}$ $\times 10^{\paramTENFOUREOFEXPONENT}$\\
$\eta_{2}$		& \paramTENFIVESOSmin $\times 10^{\paramTENFIVESOSEXPONENT}$	& \paramTENFIVESOSMCMCmin$^{+\paramTENFIVESOSMCMCplus}_{-\paramTENFIVESOSMCMCminus}$ $\times 10^{\paramTENFIVESOSEXPONENT}$	& \paramTENFIVEFTFmin $\times 10^{\paramTENFIVEFTFEXPONENT}$	& \paramTENFIVEFTFMCMCmin$^{+\paramTENFIVEFTFMCMCplus}_{-\paramTENFIVEFTFMCMCminus}$ $\times 10^{\paramTENFIVEFTFEXPONENT}$	& \paramTENFIVEEOFmin $\times 10^{\paramTENFIVEEOFEXPONENT}$	& \paramTENFIVEEOFMCMCmin$^{+\paramTENFIVEEOFMCMCplus}_{-\paramTENFIVEEOFMCMCminus}$ $\times 10^{\paramTENFIVEEOFEXPONENT}$\\
$A_{01}$		& \paramTENSEVENSOSmin 						& \paramTENSEVENSOSMCMCmin$^{+\paramTENSEVENSOSMCMCplus}_{-\paramTENSEVENSOSMCMCminus}$ 			 		& \paramTENSEVENFTFmin						& \paramTENSEVENFTFMCMCmin$^{+\paramTENSEVENFTFMCMCplus}_{-\paramTENSEVENFTFMCMCminus}$ 					& \paramTENSEVENEOFmin						& \paramTENSEVENEOFMCMCmin$^{+\paramTENSEVENEOFMCMCplus}_{-\paramTENSEVENEOFMCMCminus}$\\
$A_{02}$		& \paramTENEIGHTSOSmin 						& \paramTENEIGHTSOSMCMCmin$^{+\paramTENEIGHTSOSMCMCplus}_{-\paramTENEIGHTSOSMCMCminus}$  					& \paramTENEIGHTFTFmin						& \paramTENEIGHTFTFMCMCmin$^{+\paramTENEIGHTFTFMCMCplus}_{-\paramTENEIGHTFTFMCMCminus}$ 					& \paramTENEIGHTEOFmin						& \paramTENEIGHTEOFMCMCmin$^{+\paramTENEIGHTEOFMCMCplus}_{-\paramTENEIGHTEOFMCMCminus}$\\
\enddata
\label{CHROMATIC}
\tablenotetext{a}{As discussed in $\S$\ref{SecLargeNumberOfDataPoints} the formal errors have been scaled up by a factor of 10, as the uncertainties
presented here are believed to be a more accurate determination of the uncertainty in the fit parameters.}
\end{deluxetable*}

\begin{deluxetable}{lrrrr}[tbp]
\tabletypesize{\footnotesize}
\tablecaption{Relative Position Angles of the Primary, Secondary and the Composite Disk \label{TABLE xx}}
\tablewidth{0pt}
\tablehead{
\colhead{Extension} & \colhead{F435W} & \colhead{F606W} & \colhead{F814W} & \colhead{Average}
}
\startdata
Primary NE & $0.454$ & $0.478$ & $0.413$ & $0.448\pm0.033$ \\
Primary SW & $0.460$ & $0.519$ & $0.430$ & $0.470\pm0.045$ \\
Secondary NE & $2.567$ & $2.400$ & $2.929$ & $2.632\pm0.270$ \\
Secondary SW & $5.342$ & $5.006$ & $4.003$ & $4.784\pm0.697$ \\
Composite NE & $0.643$ & $0.565$ & $0.446$ & $0.551\pm0.099$ \\
Composite SW & $0.880$ & $0.782$ & $0.707$ & $0.789\pm0.086$ \\
\enddata
\tablenotetext{*}{All values given in degrees and measured counterclockwise from the major projection axis. The fits are obtained from regions used by \citet{gol06}, namely 80-250\,AU for the primary and the composite and 80-130\,AU for the secondary.}
\label{SPINECPSTAB}
\end{deluxetable}

\subsection{Phase Function} 
A topic of importance that has largely been ignored until now is the appropriateness of different phase functions to model
the scattering of light in the $\beta$ Pictoris disk. Since an observed brightness contour is a function of both
the distribution of scatterers as well as the phase function that describes how these particles scatter light, a discussion of the
latter is necessary. 
The zodiacal scattering profile (Fig. \ref{SCAT}) was initially chosen for two reasons: 1. previous successful modeling with this function
(eg. \citealt{kal95}) and 2. its empirical formula does not require any additional parameters. As it was already noted, using this profile has allowed us to obtain a very low reduced $\chi^2=1.18$. 
However, the question remains: how valid is this choice? One way to answer this question
is to employ other phase functions.
In section 2.2 we introduced the Henyey-Greenstein function along with the zodiacal scattering profile. The former is particularly interesting because it can be modified through the scattering
asymmetry parameter $g$ to produce many profiles, several of which are shown in Fig. \ref{SCAT}. To obtain the best fit, we add the scattering  parameter to the 18 parameters describing dust distribution with two disks and run the MCMC algorithm until the different chains converge.
Intriguingly, the 20 parameter fit does not offer noticable improvements in the reduced $\chi^2$
($\sim$1.2 for both 18 and 20 parameters). 
What's more, the values of the scattering asymmetry profile favored by the parameter search
($g_1=0.64$ and $g_2=0.85$ for the two disks)
produce a forward scattering profile that largely resembles the zodiacal scattering function;
furthermore the uncertainty in $g_1=0.64$ and $g_2=0.85$ are small (see Table \ref{NineEighteenTwenty}).
The disagreement lies in the slightly higher amount of back scatter produced by the latter, although this is not significant since most of the light comes from forward scattering.
In Table \ref{NineEighteenTwenty} we present the results of our analysis of the two-disk run that allowed for varying scattering asymmetry parameters.

\section{Discussion}

\subsection{Morphology of the Composite Disk and its Components}

\begin{figure*}[htbp]
\begin{center}
\includegraphics[angle=270,scale=0.7]{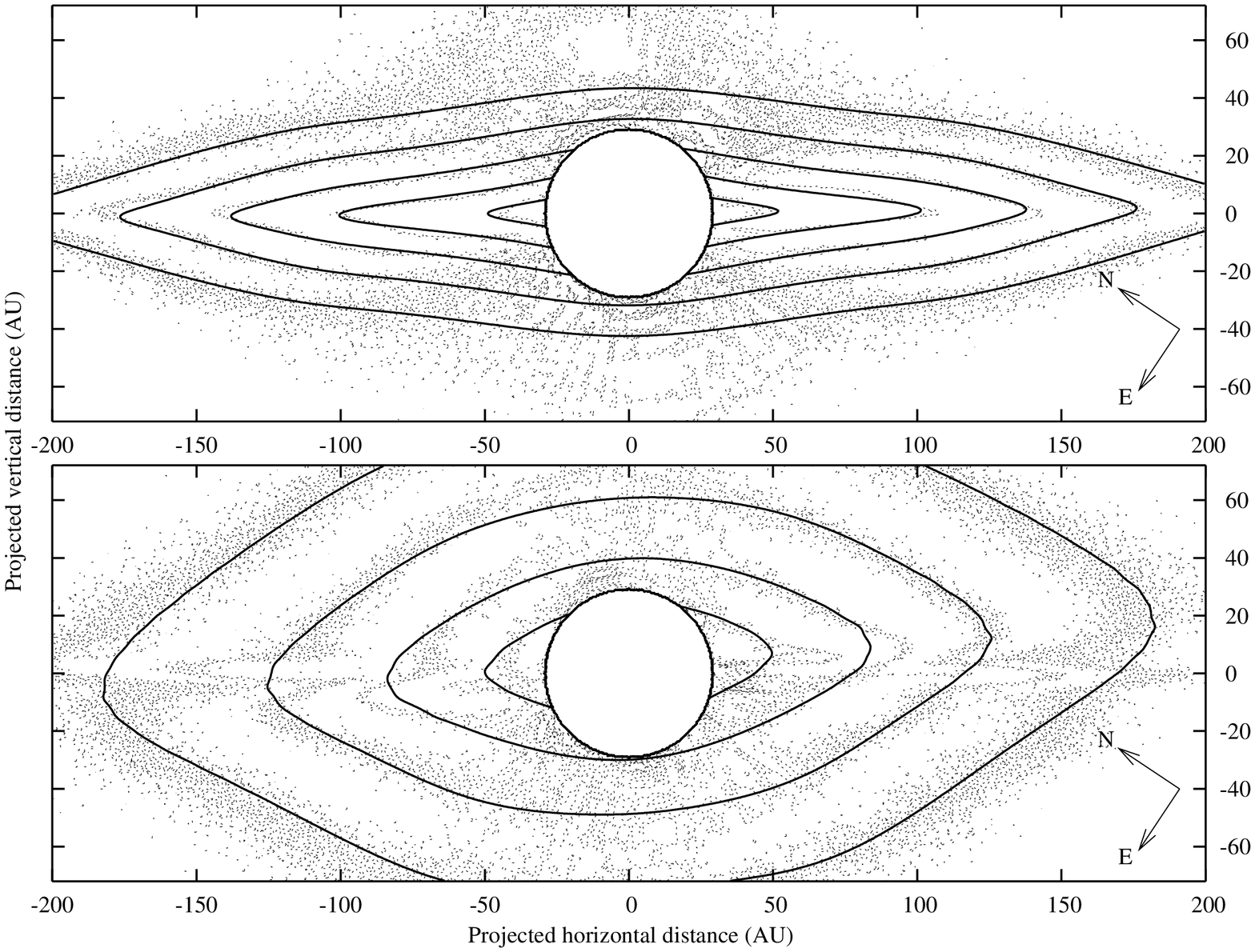}
\end{center}
\caption{Isophotal maps of light scattered by the primary (top panel) and the secondary (bottom panel). Solid lines are the isophotes of the models while the grey regions underneath represent residuals following the removal of the secondary (top panel) and the primary (bottom panel) fits. The brightness isophotes are quoted relative to $\beta Pic$, with the isophotes decreasing from $1\times10^{-7}$ $L_{disk}/L_{\beta Pic}$ in the case of the primary and from $3.162\times10^{-7}$ $L_{disk}/L_{\beta Pic}$ in the case of the secondary. In both panels the spacing of the isophotes is equal to 0.5 in log base 10 units. The empty circular region in the centre of the image represents the coronographic spot covering this area. The most important features of these fits include 1. extended secondary which contributes more light than the primary at larger radii and scale heights, 2. more widely spaced isophotes to the NW than SE of the secondary demonstrating signficant inclination and 3. nearly edge-on ($i_1=0.0$) primary.}
\label{SEPDISKS}
\end{figure*}

\begin{figure}[tbp]
\begin{center}
\includegraphics[angle=270,scale=0.7]{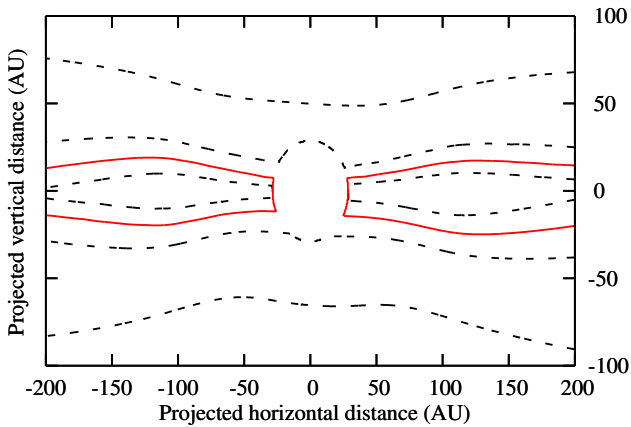}
\end{center}
\caption{Ratio of the projected light scattered by the two disks, demonstrating that 
along the midplane light from the primary dominates, while although at larger scale heights the 
secondary becomes dominant. The solid line outlines the region inside of which the primary's light dominates.
The dashed line inside the sold lines denotes
the isophote where the primary's contribution is twice that of the secondary.
The dashed lines outside the region represent isophotes where the secondary's contribution is twice and ten times greater than the primary's,
the latter being the isophote at larger scale height.
}
\label{RATIOOFSCATLIGHT}
\end{figure}

The overall shape of the brightness contours produced by the two-disk model follows that of the light scattered by $\beta$ Pic's dust. This is evident from both the isophotal maps (Figure \ref{TOT9AND18}) and low $\chi^2\sim1.2$ fits for the F606W filter images. Similar maps and reduced $\chi^2$ are recorded in other filters as well. The brightness contours of the two components (the primary and the secondary) is presented in Figure \ref{SEPDISKS} and the relative contribution of each component to the
total scattered light is shown in Figure \ref{RATIOOFSCATLIGHT}.

The most striking feature of these fits is that the secondary contributes more light at larger radii and scale heights than the primary. The immediate implication is that the dust content of the secondary disk is not localized to the inner regions of the system and that it extends at least out to $\sim250\,\mathrm{AU}$. 
We noted earlier that \citet{gol06} found a steeper power law fit along the spine of the primary beyond $\sim120\,\mathrm{AU}$ than the secondary. Their result suggests that the secondary will extend further out, at least radially, unless the fit is not appropriate beyond $\sim250\,\mathrm{AU}$ and is in line with our conclusions. It is also interesting that to describe the brightness profile along the midplane of the composite, \citet{gol06} use 4 power laws with power law breaks at 69, 117 and 193\,AU. Our two-disk model offers an explanation for the locations of the breaks. The 69 and 117 AU breaks correspond closely to the midplane profile breaks of the secondary ($71\,\mathrm{AU}$) and the primary ($113\,\mathrm{AU}$). The last power law break according to our model is a result of primary's dust density falling to levels equal to the secondary's density along the line of sight and the contour transitioning to follow the shape of the secondary. 

In order to test for possible disagreements with previous ground base observations of the disk which extend out to 800AU (e.g. \citealt{kal95}) 
and do not show two separate components, we extend our model to the same radius. Due to the large scale height of the secondary disk in our model, 
we find that the composite still appears as a single disk at large radial separations (out to $\sim$800AU). Furthermore, the primary  
makes a significant contribution to the total light profile along the spine, having the effect of aligning the composite's spine with its own. Thus, 
the composite does not show an inclination to the primary at 800\,AU. 

The impact of such an extended secondary on dust dynamics in the system is discussed in section 4.3.

\begin{figure}[tbp]
\begin{center}
\includegraphics[angle=270,scale=0.7]{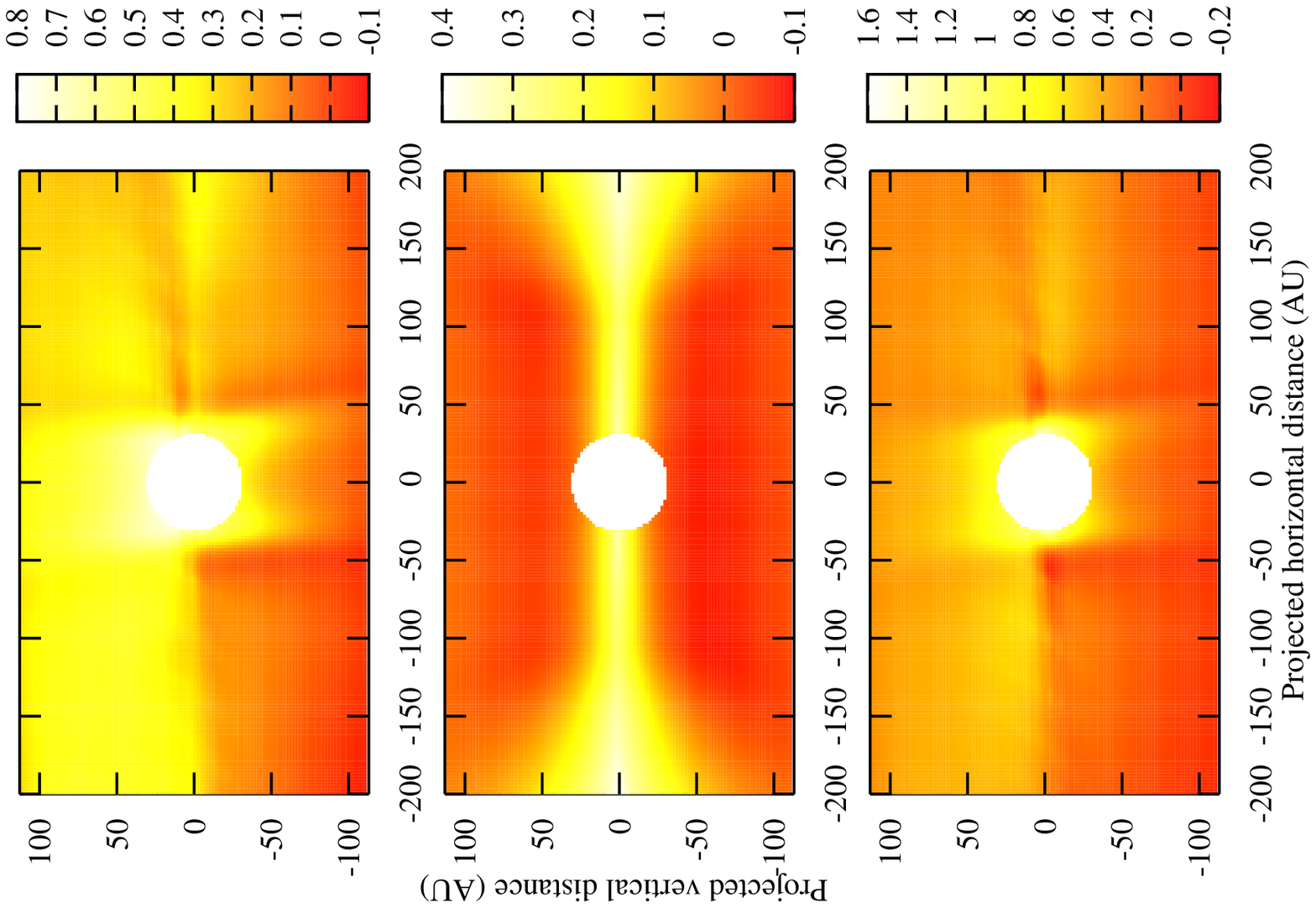}
\end{center}
\caption{Ratios of light scattered by the composite (top panel), primary (middle panel) and secondary (bottom panel) disks at 435 and 814\,nm. The red region shows the highest relative contribution in the 814\,nm relative to the 435\,nm filter. The ratios are presented in log base 2 units. From the presented ratios, we conclude that the color asymmetries along the projected major and minor axis are due to an inclined secondary disk.}
\label{CHROMRATIOS}
\end{figure}

Figure \ref{TOT9AND18} also reveals that the brightness isophotes of the two-disk (composite) model are spaced more widely along the northwestern 
semiminor axis, as can be seen from the greater extension of the isophotes in the northwest direction. \citet{gol06} attribute this spacing to the northwest side of the disk being tipped nearer to earth. All of our two-disk models agree 
that the spacing is a result of the secondary's inclination to the line of sight ($i_2=6.0\pm1.0\degr$ vs $i_1=0.0\pm0.1\degr$ for the primary which 
is almost entirely edge on). This is highlighted in Figure \ref{SEPDISKS} which maps the isophotes of the scattered light of the primary and the 
secondary showing that the secondary's isophotes are more widely spaced in the NW than the SE, where as the primary fails to shows this asymmetry. 

It is also interesting to compare brightness contours of the primary and the secondary at different wavelengths. In Figure \ref{CHROMRATIOS} we present color ratios of the F814W and F435W filters for the primary and the secondary. \citet{gol06} noted that the composite disk ratio shows asymmetry along both the major and the minor projection axis. The later of these was attributed to dust grains scattering blue light more isotropically than red light. Our model re-creates this asymmetry (Figure \ref{CHROMRATIOS}-top panel). The ratios of the primary at different filters (Figure \ref{CHROMRATIOS}-middle panel) do not show this asymmetry, which is expected from an axisymmetric edge-on disk. Our models suggest that it is the secondary which is responsible for the mentioned minor-axis asymmetry (which is expected considering its much larger inclination to the line-of-sight as well as the primary, which was used to define the major and the minor axis). 

Lastly, we comment on the residuals obtained from the subtraction of the composite from the original image (Figure \ref{TOT9AND18}). As has already 
been noted, the residual is symmetric resembling rings observed by \citet{wah03} in the mid-IR. The radii of these remnants are $(65-85)\,\mathrm{AU}$ 
and $(110-150)\,\mathrm{AU}$, with the inner ring-like structure having a similar radius and orientation as the most outer ring detected by \citet{wah03} 
(R=$82\pm2\,\mathrm{AU}$, i=$-2\pm2\degr$) and coinciding with the HST warp detected by \citet{hea00}. Since this structure is detected in the visual and 
the mid-IR regimes, we do not doubt its existence. Furthermore, if the inner ``ring'' is indeed a structure separate from the two disks, it is not 
surprising that our simple axisymmetric descriptions fail to account for its presence. Unlike the inner ``ring'', the outer structure has not been 
previously observed. This raises a possibility that it is a byproduct of the fitting procedure. For example, the two-disk fit might be overcompensating 
for the inner warp by slightly inclining the secondary disk and increasing the overlapping region. Our models fail to identify the 52\,AU ring, confirming
the absence reported by \citet{gol06}.

\subsection{Asymmetries}

High resolution images of dust surrounding $\beta$ Pictoris obtained by \citet{gol06}, with their excellent morphological detail
and photometric accuracy, impose significant constraints on models. In addition, ratios of ACS images before and after deconvolution show that dust around $\beta$-pic is distributed in two separate disks inclined at $\sim5\degr$ to each other. A good model should therefore be able to reproduce the morphology with two nearly edge-on disks inclinded by a small angle. As shown already, the low reduced $\chi^2$ and the brightness isophotes shown in Figure \ref{TOT9AND18} demonstrate the success of the MCMC algorithm we have employed to sweep through the applicable parameter space. The small inclination of the primary to the secondary ($3.2\pm1.3\degr$) and the line of sight ($\mathrm{i}_1=0.0\pm0.1\degr$) adds additional credibility to our fits.
However, no model of $\beta$ Pic's dust can be said to be successful unless it recreates at least several asymmetries mentioned in the literature, such as the wing tilt and the butterfly asymmetry. Below we discuss the known asymmetries and the degree to which our model recreates them.

\bf{Butterfly Asymmetry:} \normalfont The asymmetry first identified by \citet{kal95} refers to the asymmetric curvature of the isophotes across the spine of the composite disk and inversion of this asymmetry across the minor axis. Figure \ref{TOT9AND18} confirms that the two-disk model does a good job of re-creating this feature. Also, since the primary disk is edge-on and nearly identically aligned with the major-projection axis, it is the extended secondary that is responsible for this feature. 
  
\bf{Wing-Tilt Asymmetry:} \normalfont \citet{kal95} noted that the two extensions of the composite disk are inclined to each other by $1.3\degr$ and that the linear fits for their midplanes do not intersect at $\beta$ Pic. This asymmetry was named the wing-tilt asymmetry and is attributed to forward scattering from an inclined disk. It is also present in the \citet{gol06} images, where the inclination of the primary's components is determined to be $0.9\degr$ while that of the secondary's is $0.3\degr$. Also, they note that the primary's linear fits intersect at a point almost coincident with $\beta$ Pic, which they note as a point of disagreement. 

Our fits also reproduce this asymmetry (Table \ref{SPINECPSTAB}). The inclination of the primary's extensions ($\sim0.1\degr$), however, is much less than that reported by \citet{gol06}, but is consistent with a disk that is edge-on as our model suggests the primary is. 
The linear fits of the disk's spine are nearly centered on the star (Figure \ref{SPINECPS})
in agreement with \citet{gol06}. The secondary's extensions on the other hand are inclined at $\sim2.2\degr$ due to the inclination of this component to the line of sight ($i_2=6.0\pm1.0\degr$). Their fits also intersect close to $\beta$ Pictoris. The composite is inclined at $\sim0.3\degr$, which is less than the inclinatinon of the composite determined by \citet{kal95}. However, the two extensions do not meet at the star, with the NE component intersecting the SW component $\sim50\,\mathrm{AU}$ away from $\beta$ Pic along the SW extension. The fact that the model primary and composite do not agree on the point of intersection is offered as a possible reason for the disagreement between \citet{kal95} and \citet{gol06}. Our model suggests that it is the combination of the primary and the secondary that together creates the mentioned offset.

\begin{figure}[tbp]
\begin{center}
\includegraphics[angle=270,scale=0.7]{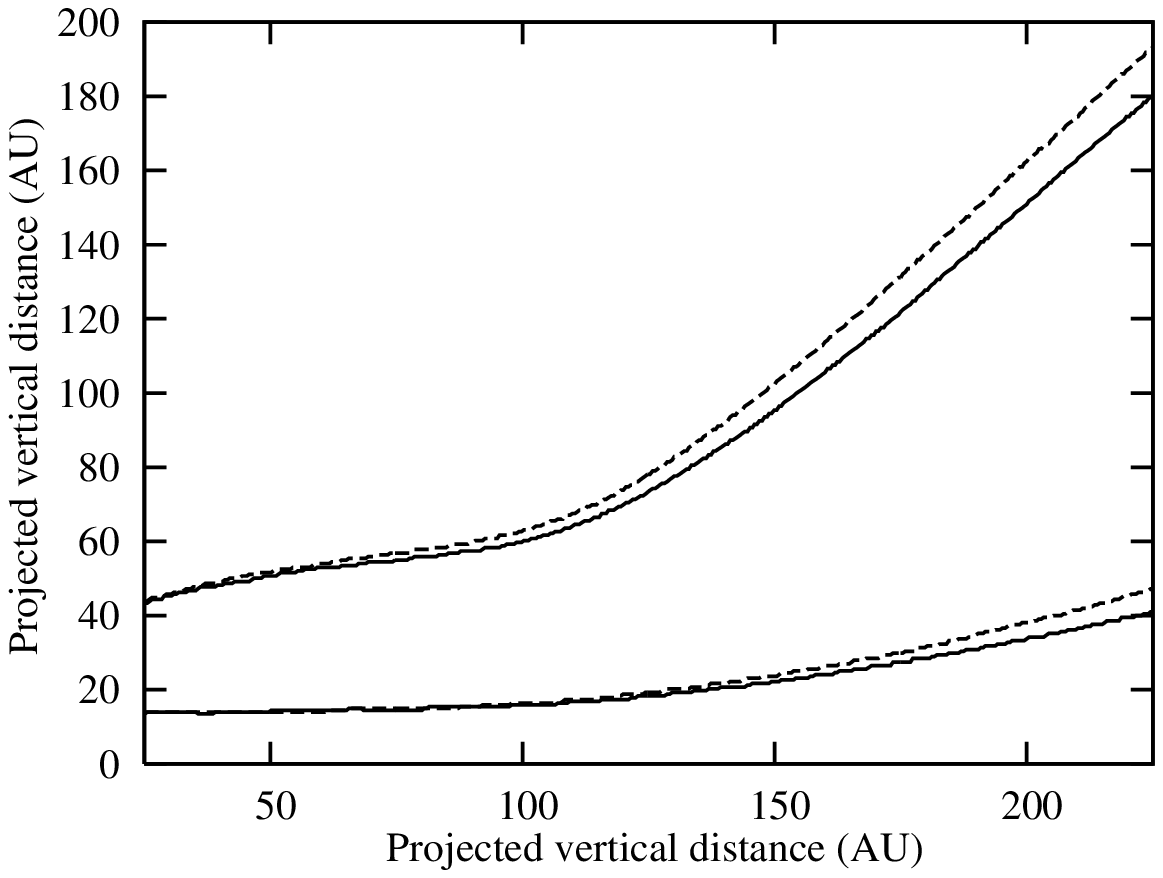}
\end{center}
\caption{FWHM (lower pair) and the FW0.1M  (upper pair) of our model
for the NE (solid lines) and SW (dashed lines).
The wider SW extension demonstrates that our model reproduces
the width asymmetry first noted by \citet{kal95}.}
\label{HEIGHT}
\end{figure}

\bf{Width Asymmetry:} \normalfont \citet{gol06} find that the FWHM and the full-width-0.1-max (FW0.1M) of the two extensions of the composite diverge beyond $\sim190\,\mathrm{AU}$ and $\sim150\,\mathrm{AU}$ respectively. This asymmetry, known as the width asymmetry, is present in our fits as well, although to a smaller degree. Figure \ref{HEIGHT} shows the width of the NE and SW components of the composite.
It is evident that the SW component's FWHM is slightly wider than that of the NE component, although not by a significant amount.
Out to $\sim200\,\mathrm{AU}$ the two extensions FWHM and FW0.1M 
almost overlap, as is the case with the \citet{gol06} measurements, but after this point our model
disagrees significantly as
the \citet{gol06} measurements show much more asymmetric behaviour.
The FW0.1M shows more asymmetric behaviour than the FWHM, as is the case with the \citet{gol06} measurements.

\begin{figure}[tbp]
\begin{center}
\includegraphics[angle=270,scale=0.7]{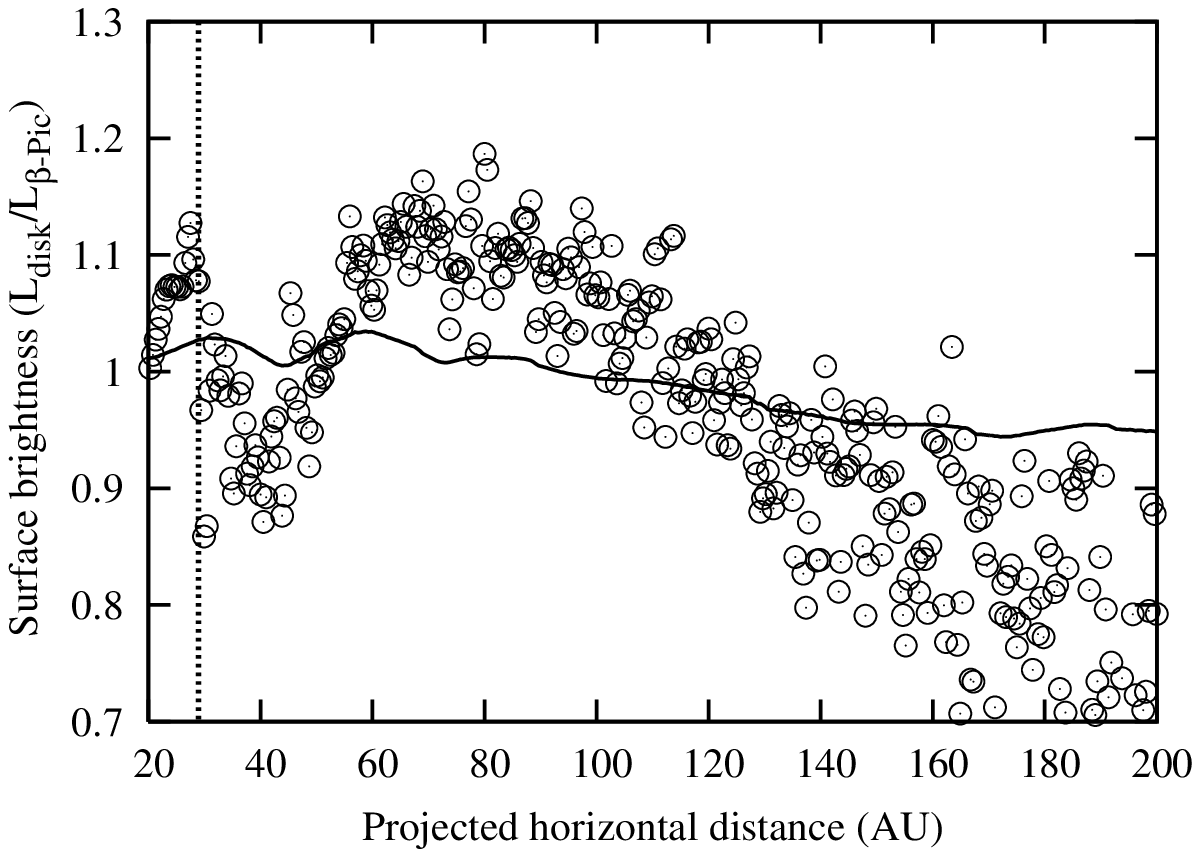}
\end{center}
\caption{The solid line represents the ratio of the brightness profile along the midplane of the SW extension
divided by the corresponding brightness profile of the NE extension of our best-fit 18-parameter model to the 606 ACS data.
The same ratio is shown for the 606 ACS scattered light data (open circles).
All open circles represent a 3 pixel average taken along the minor projection axis (the y-axis of Figure 4).
The model profile along the spines of the two extensions shows asymmetric behaviour,
but the degree of this asymmetry is far less than that observed from the scattered light.
The dotted vertical line at 30\,AU represents the extent of the coronographic mask.}
\label{RADRATIO}
\end{figure}

\bf{Surface Brightness Asymmetry:} \normalfont Differences in the brightness profile along the spine of the NE and the SW extensions of the composite are presented in Figure \ref{RADRATIO}. Although our model shows asymmetry between the profiles of the two extensions, it does this to a much lesser degree than the scattered light.

\bf{Radial Extent Asymmetry:} \normalfont 
Although the radial extent asymmetry could result from
the surface brightness asymmetry, \citet{kal95} classify it separately to draw attention to the possibility that the SW extension is physically truncated. This asymmetry, however, becomes evident only beyond $\sim400\,\mathrm{AU}$ and thus we are unable to comment on its true nature.

\normalfont

\subsection{Dust Disk Dynamics and Origin}

In previous sections we presented a case for two axisymmetric disks around $\beta$ Pic. Here we explore the dynamical consequences of such a pair. 
A quantity of interest is the mean collision time since it reveals the timescales on which the disks are stable, and also on which they have to be replenished. 
It can be estimated most easily by using our knowledge about the total stellar flux absorbed by dust and re-radiated in infrared, quantified as 
the ratio of IR excess to star's luminosity $f_d= L_d/L_*=2.4\times10^{-3}$ (e.g., \citealt{art97}).
If the absorbing area of grains in a small volume of space is denoted as $dA_{grain}$, 
then we can normalize our model-derived 
 vertical optical depth functions $\tau (r)$ using 
\begin{equation}
	f_d= \int{\frac{\frac{L_*}{4\pi r^2}\,dA_{grain}}{L_*}} \\ =
	\int{\frac{k\tau (r) 2 \pi r dr}{4\pi r^2}}=\frac{1}{2}\int{\frac{k\tau (r)  dr}{r}}.
\end{equation}
We obtain the normalization constants $k$ for each disk separately, 
utilizing the ratio of scattered light in the two disks ($L_{d,sec}/L_{d,prim}=1.23$ from our models) and the fact that $f_d$ is due to the sum of the two disks. 
The normalization factors $k_{\mathrm{primary}}$ and $k_{\mathrm{secondary}}$ 
for the two disks can then be used separately or as a sum, to discuss the behavior of
individual disks or their sum. In fact, as there is a large degree of overlap between the disks,
the second perspective is more realistic. 
We estimate the mean collision time ($t_c$) following \citep{art97} as
\begin{equation}
	t_c=\frac{P(r)}{12\tau (r)}
\end{equation}
where
$P(r)$ is the Keplerian period of a particle at radius r. The factor 12 in denominator accounts for the motion parallel to the disk and the fact that the cross section for collision between two particles of comparable radius is up to 4 times bigger than the geometrical cross section of either one. 

\begin{figure}[tbp]
\begin{center}
\includegraphics[angle=270,width=3.5in]{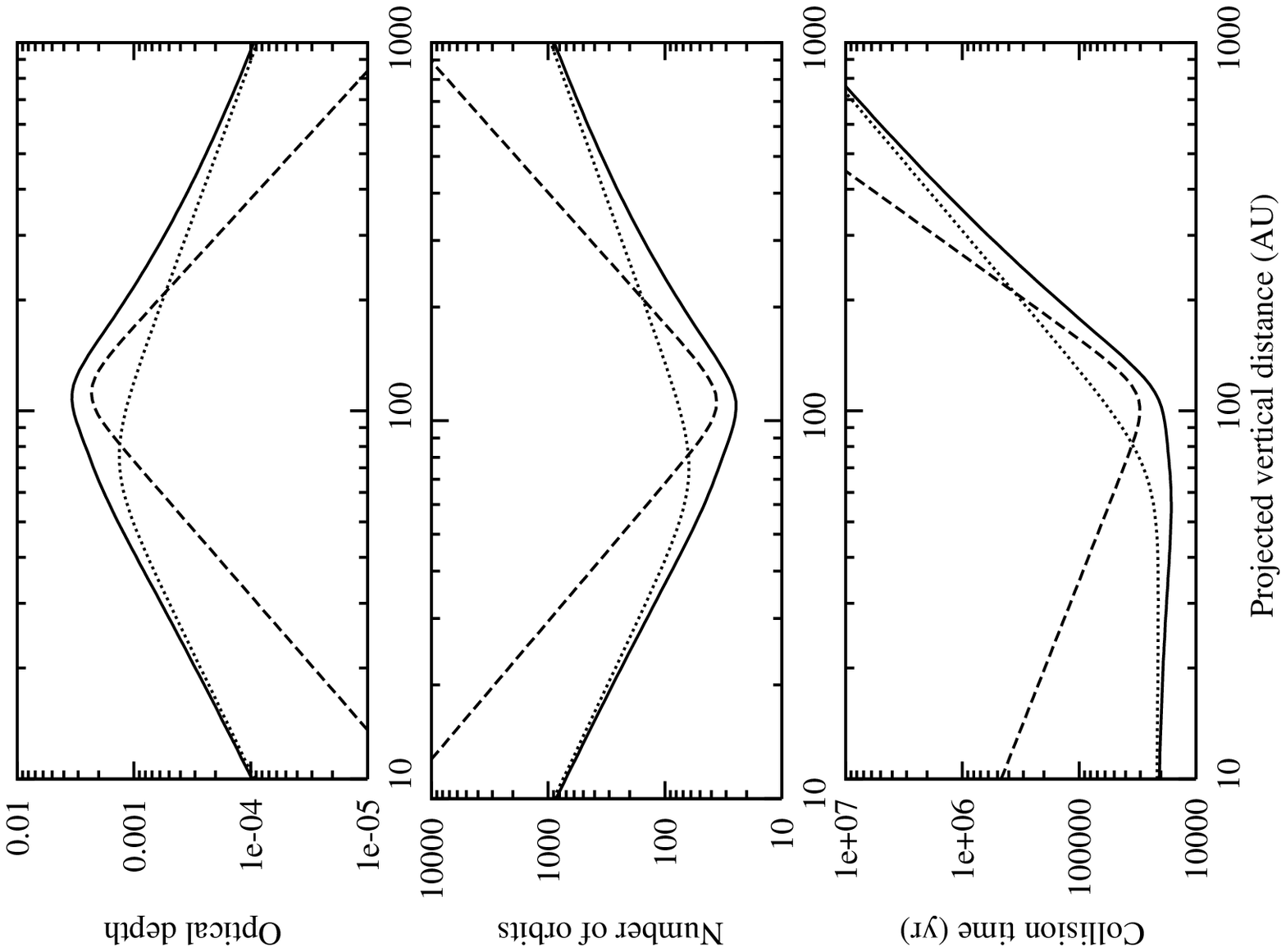}
\end{center}
\caption{The optical depths and mean collisional times for the two disks (assuming they are completely physically separated; dashed lines for the primary, dotted lins for the secondary) as well as the conservative estimate (assuming that the two disks overlap completely; solid lines).  Due to
the uncertainty in our model of the inner radial index ($\alpha_2$), the collisional times for the inner regions of the secondary disk ($<\sim60$\,AU) are less reliable.}

\label{COLLTIMES}
\end{figure}

In Figure \ref{COLLTIMES} we present the conservative number of orbits (assuming the disks completely overlap) 
a particle can make before colliding with another particle as a function of distance, and the corresponding mean collisional time scale.

Since the lifetime of dust in the densest parts of the disks is of order 
$10^{4}$ to $10^{5}$ yr, and hence much less than that of the host star (which is at least 20 Myr old), our calculations suggest both disks have planetesimals embedded in them capable of replenishing the lost dust. In other words, none of the disks represent primordial protoplanetary disks. Rather, 
they are replenished disk(s) owing their existence to a substantial mass reservoir, equivalent to 
a planetary system comparable to our own \citep{art97}. 

Geometrically, the two disks largely overlap and some regions contain comparable areas of dust. Physically, this implies mutual influence through collisions. Since the optical depths of the disks are much smaller than unity, particles in the two disks can coexist for as many as 20-10000 of their orbits. Similarly, the smaller optical depth of the larger parent bodies assures their stability for a large number of orbits in the overlapping disks. 
The inferred presence of two disks suggests that $\beta$ Pic has two distinct populations of parent bodies (planetesimals) in a slightly inclined configuration. 

As mentioned in the introduction, the characterization of the distribution of dust around a star 
is but a first step toward the derivation of the mass and spatial 
distribution of larger, unseen bodies (planetesimals), as well as any perturbing planets; other effects such as dust-dust and any gas-dust interaction in the presence of stellar radiation pressure can spatially displace the fine dust from the place of their origin (the parent bodies). However, some implications of our models are already apparent.

The fact that the secondary disk contains a significant fraction of the dust in the system
means that its origin cannot be ascribed to a minor event such as a single comet or an asteroid disruption. In order to produce comparable amounts of dust in the two disks, they must either have a common massive reservoir of parent bodies to draw from or two such reservoirs. Only comprehensive modelling based on high-quality data can clarify the situation since the intermediate-size bodies such as km-sized planetesimals are not amenable to direct detection, possessing neither enough gravitational influence nor large emitting surface area.  

The vertical profile we obtained is $\rho \sim e^{-\left(| z|/W(r)\right)^p}$, with $p=p_1\approx 0.83$ for the
first disk and  $p=p_2\approx 0.46$ for the second. Thus, neither disk 
has a vertical scattering profile resembling a disk with a  vertical isothermal structure ($p=2$,
or Gaussian vertical profile). The two disks, especially the second one, have super-exponential drop-offs ($p < 1$)
 and therefore extended,
slowly diminishing outer wings of the vertical structure, with a sharply pointed inner core. 
This could be a sign that the inner core of the profile, referred to as the spine in this paper, 
is composed of bigger particles than the profile wings; the mass separation might arise due to energy equipartition leading to a smaller velocity dispersion of the larger particles in the disk. In addition, a more fanned out secondary debris disk could be a consequence of more energetic collisions of its particles, perhaps indicating their generally smaller sizes and larger radiative modification of orbits. 

An expectation based on the the previous dynamical studies of the mobility of dust 
is that the bulk of a disk or a ring of planetesimals will be located interior to the dust disk. 
This follows from the outward action of radiation pressure exerted on dust. If so, we expect 
that most planetesimals will be found at a distances less than 50-100 AU from $\beta$ Pic.
At such distances, planetesimals can be perturbed by the proposed 8-Jupiter mass 
planet at a mean distance of 8 AU and/or other unknown planetary masses. 
Perturbing planets, in any case, must be located inside the cross-over radius ($R_{m2}$ $<$ 70 AU) of the secondary disk.

We stress here the possibility, indeed necessity, of multiple planets in $\beta$ Pic. 
A planetary system born with only one major planet would tend to stay in alignment with the orbital 
plane of this planet. In such a case, we would not expect to see two separate inclined disks. 
What could be the origin of the inclined orbit of an additional planet? 
Although specific inclination pumping mechanisms have yet to be quantified, this may provide a natural explanation
to the inclination of planetesimals and dust from the second disk. This could result
in the planet(s) themselves being on slightly inclined, mutually interacting, orbits thus contributing to the survival
of the two inclined disks.

Future observations of the inner regions of the disks would be instrumental in establishing the
presence of the putative multiple planets via their direct influence on dust. At some point, the assumption made in the present work that the disks are intrinsically axisymmetric, if misaligned, must prove inadequate. With sufficient observational data, our assumption of the same optical properties of dust in two disks may also be relaxed. Furthermore, a future comprehensive model of the system must include not only collisional dust dynamics and radiation, but also the collisional cascades spanning a large size range from planetesimals to micron-sized dust, and possibly more unusual ingredients, such as dust avalanches involving sub-micron dust. 

\section{Conclusions}

Below we  summarize the results of this work.

\begin{enumerate}

\item We modeled the dust distribution around $\beta$ Pic with two axisymmetric disks. A MCMC $\chi^2$ optimization method was used to reject disk models and phase functions that produce brightness contours different from those obtained with the ACS by \citet{gol06}. We find a small quantitative improvement in the reduced $\chi^2$ (from 1.8 to 1.2) of 
the two-disk fit compared to the single disk fit; the two-disk fit, however, displays significant improvements in its ability to trace the isophotes of the composite. The parameters listed in Table \ref{CHROMATIC} along with a zodiacal light scattering profile produce the best fits. 

\item Dust distribution in the primary disk is largely in agreement with conclusions from previous modeling attempts
(eg. \citealt{art89,kal95}). The modelled primary disk has a near-zero inclination to the line of sight. Therefore, the up-down asymmetry of the whole disk
 is not caused by the primary. The surface density of dust increases as  $\alpha_1=2.73$-th power of radius inside a
cross-over radius $R_{m1}=$113 AU and decreases at large radii with power-law index $\beta_1=-2.89$
(values from F606W filter fit). This behavior is similar to the one known from previous modelling, although the inner power law is steeper. The 
vertical disk profile has a nearly exponentiall fall-off with vertical distance ($p_1=0.85$). 
The scattered light contours from the primary disk are shown in Figure \ref{SEPDISKS}. 

\item The secondary disk model presents the first quantitative measurement of the dust distribution in this component. It suggests that the
secondary disk, with a scale-height much larger than the primary, is not radially limited to the inner system, but that it extends
at least to 250\,AU, and possibly further out. The secondary disk model
 has a larger scattering area than the primary at large distances and scale heights, 
which follows from the less-steep power laws of the radial surface density ($\beta_{2}=-1.16 > -2.89$ at distances much larger than the cross-over $R_{m2}=71$ AU) and the vertical density ($p_2=0.46 < 0.85$). 
The ratio of the total scattered light (secondary vs. primary) is roughly equal to 1.23. Since we do not model the dust size distribution, we cannot compute the dust mass ratio of the two components. It is however plausible that the secondary disk is as substantial as the primary. The larger extent and especially the larger scattering area of the secondary disk are surprising and raise many questions about the origin and dynamics of such a pair of disks. The scattered light contours from the secondary are presented in Figure \ref{SEPDISKS}. 

\item Several of the asymmetries reported by \citet{kal95} can be reproduced with our model. The butterfly asymmetry is explained by the presence of an extended secondary disk inclined by $3.2\pm1.3\degr$ to the primary. We find that it is the secondary that is inclined to the line of sight and therefore responsible for the wing-tilt asymmetry observed in the composite image. The width
 asymmetry is present, but not to the degree seen in \citet{gol06} data beyond $\sim200\,\mathrm{AU}$. The brightness asymmetry is
 also noticable both across the major and the minor projection axis. The difference across both the axes is attributed to a secondary disk
 inclined by $i_2=6.0\pm1.0\degr$ to the line of sight and to the primary which is nearly aligned with the major projection axis. 

\item In order to explore dust dynamics in a pair of disks described by the parameters obtained here, we analytically constrain the dust replenishment times to be $10^4$ yr at $\sim$100 AU. Such small timescales show that both disks have to be continuously replenished and contain planetesimals that, as eventual mass reservoir, are capable of producing the 
second-generation dust. The nature and origin of the inclined disks and planetesimals in $\beta $ Pic  should be explored via dynamical models of spatially resolved collisional cascades subject to radiation pressure effects.

\item The most plausible conjectural outcome of this modeling is that the non-coplanarity of two 
dust disks is due to the secondary dust disk being supplied by its own disk of planetesimals, which has been inclined by the gravity of a planet residing inside the cross-over radii ($r<70$ AU).
This, in turn, would be most naturally explained if there were more than one massive planets in the system, in orbits inclined by several degrees. Future modeling of unseen planetesimals of $\beta $ Pictoris, constrained by the new knowledge of dust distribution, may clarify this conjecture.
\end{enumerate}


\acknowledgements
The Natural Sciences and Engineering Research Council of Canada supported the research of B.C. and P.A.
The authors thank Sally Heap and David Golimowski for data and informative discussions on their observations, and an anonymous referee for comments on the paper.

\end{document}